\documentclass[9pt,twocolumn,twoside]{myclass}
\usepackage{comment}
\usepackage{cleveref}

\title{{\Huge Learning an optimal PSF-pair for ultra-dense 3D localization microscopy}}

\author[1,2,5]{{\Large Elias Nehme}}
\author[2,3,5]{{\Large Boris Ferdman}}
\author[2]{{\Large Lucien E. Weiss}}
\author[2]{{\Large Tal Naor}}
\author[4]{{\Large Daniel Freedman}}
\author[1]{{\Large Tomer Michaeli}}
\author[2,3,*]{{\Large Yoav Shechtman}}

\affil[1]{The Erna and Andrew Viterbi Faculty of Electrical Engineering, Technion - Israel Institute of Technology, 3200003 Haifa, Israel}
\affil[2]{Biomedical Engineering Department and Lorry I. Lokey Center for Life Sciences and Engineering, Technion - Israel Institute of Technology,  3200003 Haifa, Israel}
\affil[3]{Russel Berrie Nanotechnology Institute, Technion - Israel Institute of Technology, 3200003 Haifa, Israel}
\affil[4]{Google Research, Haifa, Israel}
\affil[5]{Equal Contribution}
\affil[*]{Corresponding author: \href{mailto:yoavsh@bm.technion.ac.il}{yoavsh@bm.technion.ac.il}}

\date{}

\begin{document}

\maketitle

\begin{abstract}

A long-standing challenge in multiple-particle-tracking is the accurate and precise 3D localization of individual particles at close proximity. One established approach for snapshot 3D imaging is point-spread-function (PSF) engineering, in which the PSF is modified to encode the axial information. However, engineered PSFs are challenging to localize at high densities due to lateral PSF overlaps. Here we suggest using multiple PSFs simultaneously to help overcome this challenge, and investigate the problem of engineering multiple PSFs for dense 3D localization. We implement our approach using a bifurcated optical system that modifies two separate PSFs, and design the PSFs using three different approaches including end-to-end learning. We demonstrate our approach experimentally by volumetric imaging of fluorescently labelled telomeres in cells.

\end{abstract}

\section{Introduction} \label{sec:intro}

In a conventional imaging system, the spatial resolution is bounded by Abbe's diffraction limit. In a high numerical aperture microscope, this corresponds to approximately half the optical wavelength, \emph{i.e}.\ \(\approx\)200 nm for visible light. For cell-imaging applications, this obscures subcellular features of interest with dimensions on the nanoscale. Since 2006, Single-Molecule Localization Microscopy (SMLM) super-resolution techniques have revolutionized biological-structure imaging by circumventing the diffraction limit, namely, using many low-density images of different sets of fluorescent emitters to generate a high-resolution reconstruction \cite{betzig2006imaging, hess2006ultra, rust2006sub}.

While biological structures are intrinsically 3D, attaining axial (z) information at super-resolution is not trivial. This is due to the standard Point Spread Function (PSF) of the microscope being approximately symmetric about the focal plane, and having only a thin axial range before the signal becomes very diffuse. Several approaches have been developed to capture 3D data in microscopy. For example, one can acquire multiple 2D datasets at different focal planes \cite{ram2008high,juette2008three,louis2020fast}, or determine the axial positions of emitters from the images themselves. The latter can be enabled by PSF engineering, where the PSF is modified to encode the desired 3D information. This is typically done by either inducing an intentional aberration in the imaging path, e.g. a cylindrical lens \cite{huang2008three} or a phase mask at the Fourier plane of the microscope using an extended optical system \cite{pavani2009three,shechtman2014optimal,backer2014extending}. Notably, while providing scan-free axial information, this approach poses a limitation on the maximum emitter densities suitable for imaging, due to the increased lateral size of the PSFs, and requires more complex image-analysis algorithms than 2D localizaiton microscopy. 

When imaging samples that are even just several microns thick, engineered PSFs spread the signal photons over a large lateral footprint relative to the in-focus PSF \cite{shechtman2015precise}. This poses a difficult localization challenge when the experimental objective of obtaining a super-resolution reconstruction necessitates that many molecules be localized in a densely labelled structure. Currently available software packages struggle to achieve good performance in this regime \cite{aristov2018zola,sage2019super}; however, recent work has shown that deep neural networks are well suited to the problem \cite{mockl2020deep}, enabling high-quality reconstruction estimations from low emitter densities \cite{ouyang2018deep,gaire2020accelerating}, and increased-density processing \cite{nehme2018deep, boyd2018deeploco,newby2018convolutional,diederich2019cellstorm, nehme2020deepstorm3d, speiser2019teaching, barth2019coupling}.\par
\begin{figure*}[ht!]
\centering
\includegraphics{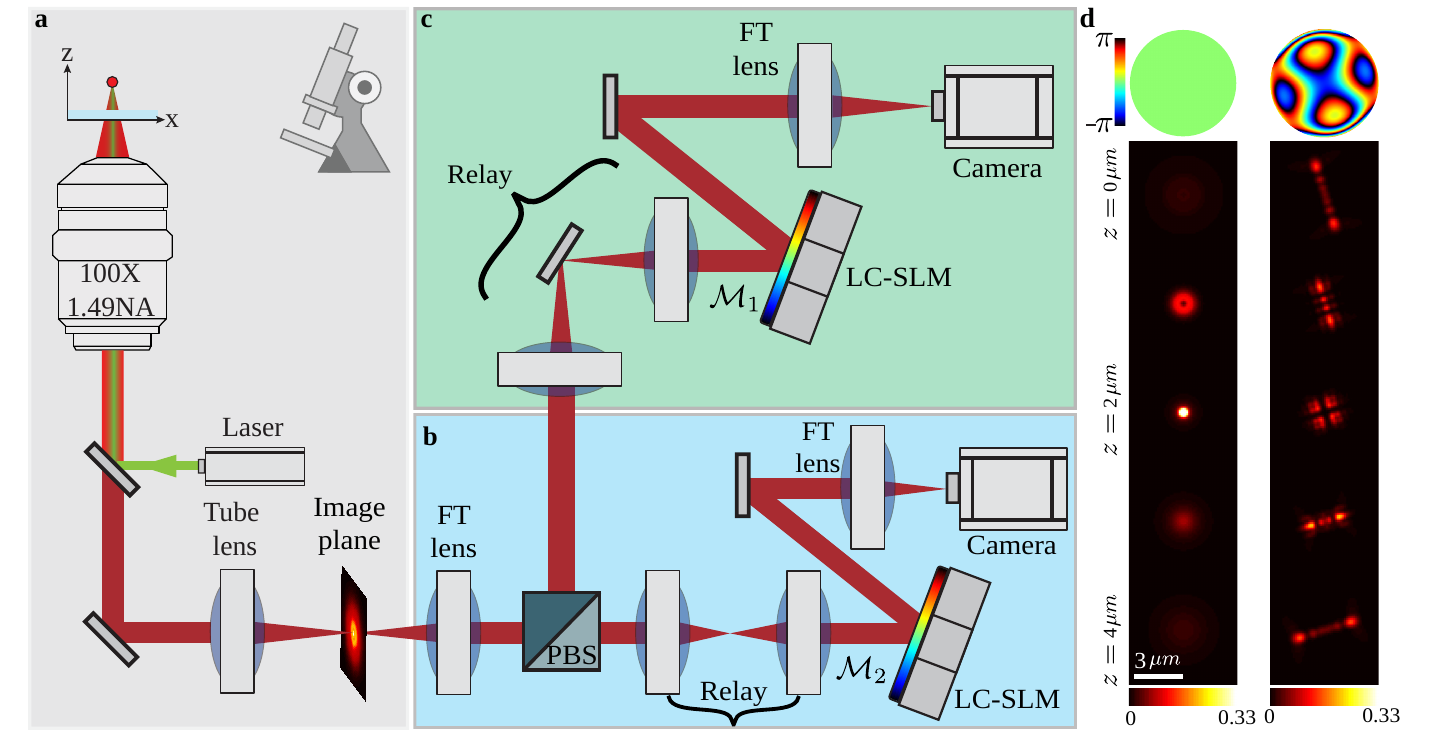}
\caption{The multi-PSF optical system. (a) A standard inverted microscope with laser illumination. (b-c) Two image planes, split by their polarization, employing two LC-SLMs placed in conjugate back focal planes to the objective lens. Each optical path can be modulated with a different phase mask (\(\mathcal{M}_1\) \& \(\mathcal{M}_2\)). (d) A comparison between the standard PSF (left) and a 4\(\mu m\) Tetrapod PSF (right).}
\label{fig:optical-stup}
\end{figure*}
Deep Learning (DL) has excelled in a variety of challenging computational-imaging problems in computer vision, computational photography, medical imaging, and microscopy \cite{barbastathis2019use,ongie2020deep}. Within the realm of computational microscopy, DL has been deployed for tasks such as cell segmentation \cite{falk2019u}, image restoration \cite{rivenson2017deep,weigert2018content,krull2019noise2void,lim2020cyclegan}, sample classification \cite{horstmeyer2017convolutional,muthumbi2019learned}, artificial labelling \cite{ounkomol2018label}, phase imaging \cite{rivenson2018phase,nguyen2018deep,xue2019reliable}, optical tomography \cite{wu2020simba}, lifetime imaging \cite{smith2019ultra}, single-molecule localization \cite{nehme2018deep,boyd2018deeploco,ouyang2018deep,newby2018convolutional,zelger2018three,zhang2018analyzing,hershko2019multicolor,diederich2019cellstorm,nehme2020deepstorm3d, speiser2019teaching,barth2019coupling,gaire2020accelerating,dardikman2020learned}, aberration correction \cite{mockl2019accurate,mockl2020accurate,saha2020practical,shajkofci2020spatially}, CryoEM \cite{gupta2020cryogan}, and more \cite{belthangady2019applications}.

An exciting recent application enabled by deep learning is the \emph{end-to-end} design of “computational cameras.” Powered by differentiable imaging models and back-propagation, end-to-end learning jointly optimizes the sensing system alongside the data-processing algorithm, thus enabling both components to work harmoniously. This approach has quickly expanded within the computational-imaging community for numerous applications in computer vision and computational photography, for example, color sensing and demosaicing \cite{chakrabarti2016learning, schwartz2018deepisp}, illumination-design through scattering media \cite{turpin2018light}, extended-depth-of-field imaging \cite{elmalem2018learned,sitzmann2018end,akpinar2019learning}, monocular depth estimation \cite{sitzmann2018end,elmalem2018learned, wu2019phasecam3d, chang2019deep}, high-dynamic-range imaging \cite{metzler2020deep,sun2020learning}, and hyper-spectral imaging \cite{dun2020learned,baek2020end}.
In computational microscopy, end-to-end learning has been utilized by our group and others to enhance various computational modalities such as sample classification \cite{horstmeyer2017convolutional,muthumbi2019learned}, single-molecule color sensing and 3D localization \cite{hershko2019multicolor,nehme2020deepstorm3d}, quantitative phase imaging \cite{kellman2019data} and multi-photon microscopy \cite{pinkard2020learned}.

Here, to address the challenge of high density 3D localization from a snapshot, we suggest the simultaneous use of multiple PSFs, as well as the method to design and implement the optimal phase masks. Specifically, we introduce a bifurcated optical system that modifies two separate PSFs with a pair of phase masks using Liquid-Crystal Spatial Light Modulators (LC-SLMs). First, we demonstrate that there is an advantage of splitting precious signal photons into two channels compared to a single PSF system even in moderately dense emitter conditions. For this task we utilize a PSF-pair that splits the 3D information into complementary channels, namely, for lateral and axial localization. To localize the emitters from the obtained pair of images we employed a convolutional neural network (CNN) architecture based on DeepSTORM3D \cite{nehme2020deepstorm3d}. Next, we revisit the problem of optimizing the information content of a single emitter in a pair of PSF measurements \cite{shechtman2014optimal}. Lastly, we implement end-to-end learning to jointly design our localization algorithm and the PSF-pair. The resulting PSFs, which we call the Nebulae PSFs, achieve unprecedented performance in localizing volumes of dense emitters in 3D. We quantify and directly compare the performance of each approach by simulation and experimentally with volumetric imaging of fluorescently labelled telomeres in fixed cells. Finally, we demonstrate continuous, scan-free, live-cell tracking of \(>\)60 telomeres in a single cell’s nucleus simultaneously with \(\approx\)30 nm 3D precision and 100 ms temporal resolution over an axial range of \(\approx\)5 \(\mu m\).

\section{Optical setup} \label{sec:opt-setup}

Dual-camera systems have been utilized in the past in microscopy for localizing single emitters in 3D \cite{lew2011corkscrew, backlund2012simultaneous, roider2014axial, min20143d}. Most recently, the use of a dual-view scheme was utilized in DAISY \cite{cabriel2019combining} to combine Astigmatism-based PSF engineering with Super-critical Angle Fluorescence (SAF) \cite{bourg2015direct} to provide a semi-isotropic 3D resolution over a \(\approx\)1 \(\mu m\) axial range. However, while these works proposed creative designs to combine the information in both channels, their objective was to enable a precise and experimentally-robust axial localization of \emph{single} emitters. In addition, the proposed PSFs were hand-crafted based on desired properties and not fully optimized. Here we use a bifurcated optical system with two detection paths for the task of precise 3d localization of multiple emitters in ultra-dense samples.

The optical system used to implement the monocular PSF-pair is presented in Fig. \ref{fig:optical-stup}. Briefly, our system is composed of an epifluorescence microscope extended with two identical detection paths. The fluorescent light emitted from the particle in the sample is split using a polarizing beam splitter into two 4\(f\) optical processing systems, each equipped with a LC-SLM placed in the Fourier plane. The LC-SLM is used to implement a phase modulation modifying the emission pattern to encode the 3D position onto the 2D captured measurements, which are then decoded jointly \textit{via} further image processing. For a list of the specific components used in our implementation see supplementary section \ref{supp-subsec-optics-comp}.


We model our system using the scalar diffraction approximation where the emitters are modeled as isotropic point sources \cite{goodman2005introduction}. Thus,  the PSFs of our system can be efficiently computed by a Fast Fourier Transform (FFT). A full description of our imaging model is provided in supplementary section \ref{supp-imaging-model}. 

Equipped with the system above, the question is what pair of PSFs is suited for the task of dense 3D localization. In the next sections we gradually answer this question.

\section{Disentangling lateral/axial information} \label{sec:xy-z}

For simplicity, we first consider the problem of designing an additional PSF while keeping the first PSF fixed to the 4\(\mu m\) Tetrapod \cite{shechtman2014optimal}, which was optimized for the sparse case in a single channel. Given that Tetrapod PSFs encode depth at the cost of a large lateral footprint, we would like the complementary PSF to be compact and help disentangle the approximate lateral positions in overlapping regions. Then, aided by this additional measurement, the overlapping Tetrapods can be decoded to recover the 3D positions. In other words, we are, broadly, separating the problem into an "axial localization" channel, encoded by the Tetrapod PSF, and a "lateral localization" channel, to be encoded by a different PSF. 

For encoding lateral information we propose the use of an Extended-Depth-of-Field (EDOF) PSF, namely, a PSF that maintains its lateral shape over extended axial ranges. However, unlike traditional EDOF designs \cite{ben2005experimental,dowski1995extended}, the desired PSF needs to be laterally-compact and signal-efficient, because it should work for very dense samples. These requirements motivated us to design a novel EDOF suited for the task.

\subsection{EDOF PSF design} \label{subsec:edof-design}

To design the desired EDOF PSF, we formulate the problem as a phase retrieval task. Specifically, given a desired axial range \(\Delta_z\) (\emph{e.g}.\ 4 \(\mu m\)s), we first generate a synthetic z-stack comprised of the approximate in-focus Airy disk PSF \(\mathcal{A}\left(x,y\right)\) at 200 nm steps. Afterwards, we use stochastic gradient descent iterations with importance sampling \cite{ferdman2020vipr} to recover the phase mask \(\mathcal{M}\) associated with this PSF. Let D be the diffraction limit for the assumed optical setup. Then our cost function for this task is given by
\begin{equation}
    \mathcal{L}_{\text{EDOF}}\left(\mathcal{M}\right) = \sum\limits_{i=1}^{N_z}{\|\left(PSF\left(x,y;\mathcal{M},z_i\right) - \mathcal{A}\left(x,y\right)\right)\cdot \mathcal{S}\left(x,y\right)\|_2^2},
\end{equation}
where \(PSF\left(x,y;\mathcal{M},z_i\right)\) is the on-axis PSF at depth \(z_i\), \(N_z\) is the number of axial slices (\(\frac{\Delta_z}{200 \text{nm}}\)), and \(\mathcal{S}\left(x,y\right)\) is a weighting term added to quickly ``squeeze" the signal photons into the diffraction limited spot, given by
\begin{equation}
    \mathcal{S}\left(x,y\right) =
    \begin{cases}
    1,& \text{if } \sqrt{x^2 + y^2}\leq \text{D}\\
    25\cdot\sqrt{x^2 + y^2},& \text{otherwise}
    \end{cases}.
\end{equation}

\begin{figure}[h!]
\centering
\includegraphics{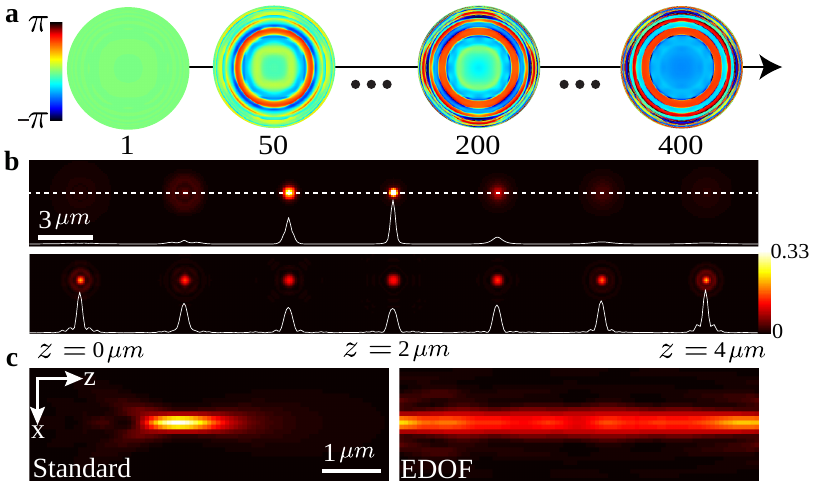}
\caption{A small-footprint EDOF mask. (a) The evolution of the EDOF phase mask optimization over 400 ietrations. (b) Comparison between the standard PSF (top) and the final EDOF PSF (bottom). (c) The XZ cross-sections of the standard (left) and EDOF (right) PSFs, respectively. The colorscale is normalized to the maximum intensity of the in-focus, unmodulated PSF.} 
\label{fig:edof-design}
\end{figure}

The resulting phase mask and PSF are presented in Fig. \ref{fig:edof-design}. This simple approach leads to a  powerful EDOF, with very high signal-efficiency and small lateral-footprint (Fig. \ref{fig:edof-design}b) compared to previous designs \cite{ben2005experimental,dowski1995extended} (see supplementary section \ref{supp-edof-design} for comparisons and implementation details). While we designed and implemented this EDOF to complement the Tetrapod information in emitter-dense regions, its potential applications extend far beyond our localization task.

Notably, recent end-to-end designs of EDOF PSFs have achieved quite compelling results \cite{elmalem2018learned,sitzmann2018end,akpinar2019learning}. In particular, the phase mask presented in \cite{akpinar2019learning} resembles the result of our approach. However, these data-driven approaches are ultimately dataset-dependant, and take hours of training to design for a new range, whereas our approach is independent of the dataset and converges in less than 2 minutes on GPU.

\subsection{Dual-view vs Single-view} \label{subsec:dual-vs-single}

In typical LC-SLM PSF engineering systems, half of the signal-photons are discarded, since the LC-SLM can only modulate polarized light. Therefore, in our system the second PSF measurement comes at no additional photon-cost, with the only caveat being the need of an additional detection path in the two-view setup. It should be noted that 4\(f\) systems that utilize a Diffractive Optical Element (DOE), instead of a LC-SLM, do not suffer from this photon loss. Yet, this comes at the cost of versatility. Now that we have designed a novel EDOF PSF for our task, we can test the hypothesis whether or not splitting the signal into two cameras is in fact beneficial compared to a DOE based system.

Since neural networks are already established to be incredibly efficient for dense localization \cite{nehme2018deep,nehme2020deepstorm3d},  we modify our previously published fully convolutional architecture  \cite{nehme2020deepstorm3d} to receive an image with two channels comprised of the two measurements. For training details and network architecture see supplementary section \ref{supp-learn-details}.
Our results in simulation (see supplementary Fig. \ref{figs:single-vs-dual-plots}) confirms that for the task of dense 3D localization, a split signal dual-view system is superior to a single measurement with a DOE, even when that measurement is sensed using an optimal end-to-end learned design \cite{nehme2020deepstorm3d}. 

\subsection{Tetrapod-EDOF experimental validation} \label{subsec:exp-tp-edof}

\begin{figure}[h!]
\centering
\includegraphics{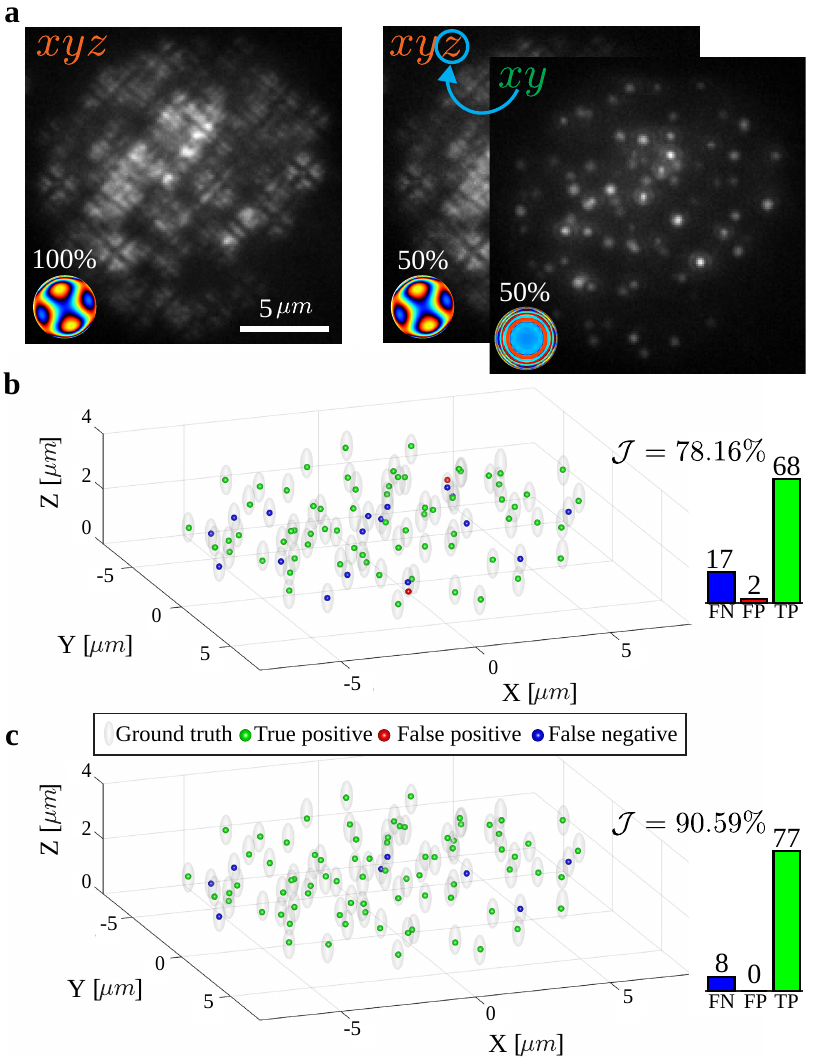}
\caption{Snapshot, dense-emitter, 3D localizations in fluorescently labelled cells. (a) A single frame recorded with a single-channel, 4 \(\mu m\) Tetrapod PSF (left) and the split-channel, dual PSF approach (right). (b-c) Localizations are plotted with the ground truth measured by axially scanning the sample with the unmodulated PSF.} 
\label{fig:tp-vs-tp-edof}
\end{figure}

Next, we validate our approach in cells. For this task, we imaged fluorescently labeled telomeres in fixed human osteosarcoma (U2OS) cells (for fixation and labeling see supplementary section \ref{supp-subsec-optics-comp}). We first chose fixed cells to enable the acquisition of a ground truth approximation \textit{via} axial scanning. The imaged cell line was hypertriploid, meaning that it has an unusually large number of telomeres (70-130), which facilitates testing our method in a dense environment. The experiment consisted of two parts: first, each cell was scanned in the axial direction using a piezo stage (100 nm steps) the 3D ground truth positions were approximated \textit{via} fitting (see supplementary section \ref{subsec:supp-calib-process}). Afterwards, we recorded 3 snapshot images: one with the Tetrapod PSF utilizing 100\% of the signal (accomplished using a longer exposure time) and two more with the signal split 50\%/50\% between the Tetrapod PSF and the EDOF PSF (Fig. \ref{fig:tp-vs-tp-edof}). In agreement with simulations, these results demonstrate that at a density of \(\approx\)0.27 \(\left[\frac{emitters}{\mu m^2}\right]\), the Tetrapod-EDOF pair is superior in localizing overlapping telomeres as measured by the Jaccard index \cite{sage2019super,nehme2020deepstorm3d}. 

While the complementary PSF-pair is effective, this way of decoupling the 3D positional information by dedicated "lateral" and "axial" channels is unlikely to be the optimal solution. For example, beyond a certain density, the axial information in the Tetrapod PSF will be occluded completely by overlapping PSFs. Having a second measurement that is solely dedicated to encode the lateral information (EDOF PSF) will not be beneficial for decoding \(z\). This motivates us to revisit the task of designing a PSF-pair for dense 3D localization. For simplicity we start with the single-emitter case, viewed from an estimation theory perspective.

\section{Optimal PSF-pair design} \label{sec:optimal-design}

\subsection{Single-emitter case} \label{subsec:crlb}

Optimal PSFs for two-channel localization of only a single emitter can be derived by minimizing the Cram\'er Rao Lower Bound (CRLB) \cite{kay1993fundamentals,ober2004localization,shechtman2014optimal}. Considering the system in Fig. \ref{fig:optical-stup}, we can jointly optimize the sensitivity of a PSF-pair with respect to a change in the 3D position of a single emitter. The CRLB then defines the lower bound on the precision of unbiased estimation of the 3D position from a noisy-PSF pair. Unlike the original Tetrapod optimization \cite{shechtman2014optimal}, here we employed a pixel-wise approach to explore aberrations not spanned by low-order Zernike polynomials. For a full derivation of the CRLB and the optimization objective see supplementary section \ref{supp-crlb-opt}.

\begin{figure}[ht!]
\centering
\includegraphics{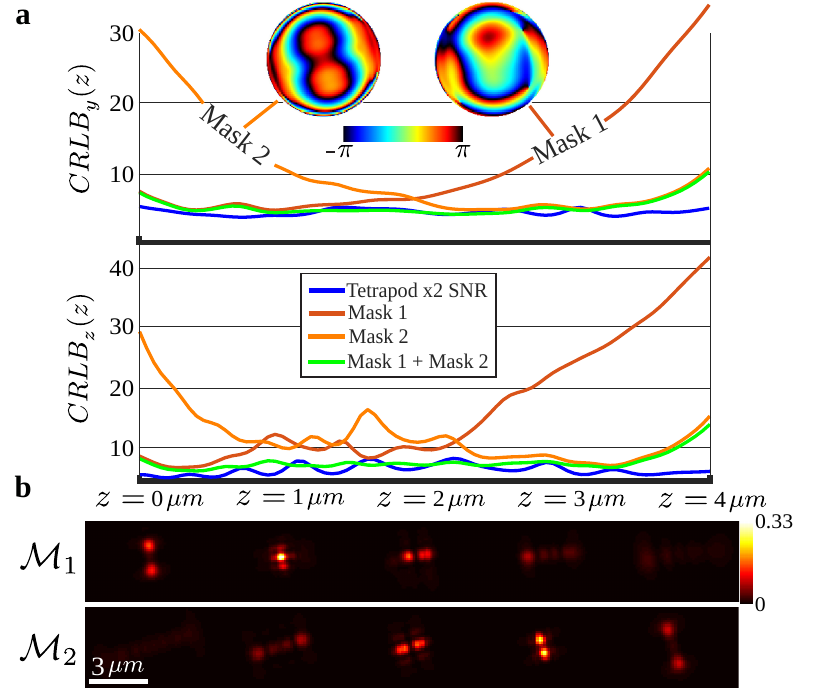}
\caption{CRLB-optimized PSF pair. (a) Two phase masks were generated by CRLB optimization, namely by estimating the 3D position of a single emitter from a pair of images. Interestingly, each channel encodes a complementary part of the axial range. These PSFs have a smaller lateral footprint than the similar z-range 4 \(\mu m\) Tetrapod. The colorbar is normalized to the in-focus unmodulated PSF of the system. (b) The estimated CRLB lateral (upper) and axial (lower) precision as a function of emitter depth of each PSF separately (red and orange), and after combining both channels (green), as well as the single-channel PSF Tetrapod (blue).}
\label{fig:crlb-2psfs}
\end{figure}

The CRLB-optimized PSF pair is given in Fig. \ref{fig:crlb-2psfs}. Notably, the CRLB of the PSF-pair is similar to the CRLB of a 4 \(\mu m\) Tetrapod PSF with twice the signal. Therefore, as can be expected, splitting the information does not improve precision in the single-emitter case, suggesting that a two-channel system is not justified for sparse localization.

The resulting PSF-pair combines the concept of bi-plane imaging and PSF engineering in an elegant way to encode the 3D position in two measurements. Simulation results show that this PSF-pair outperforms the Tetrapod-EDOF pair described earlier (see supplementary sections \ref{supp-add-sim} and \ref{supp-add-exp}); however, previous work demonstrates that end-to-end designs using deep neural networks can lead to superior performance \cite{nehme2020deepstorm3d}, and this is the path we describe next.

\begin{figure*}[ht!]
\centering
\includegraphics{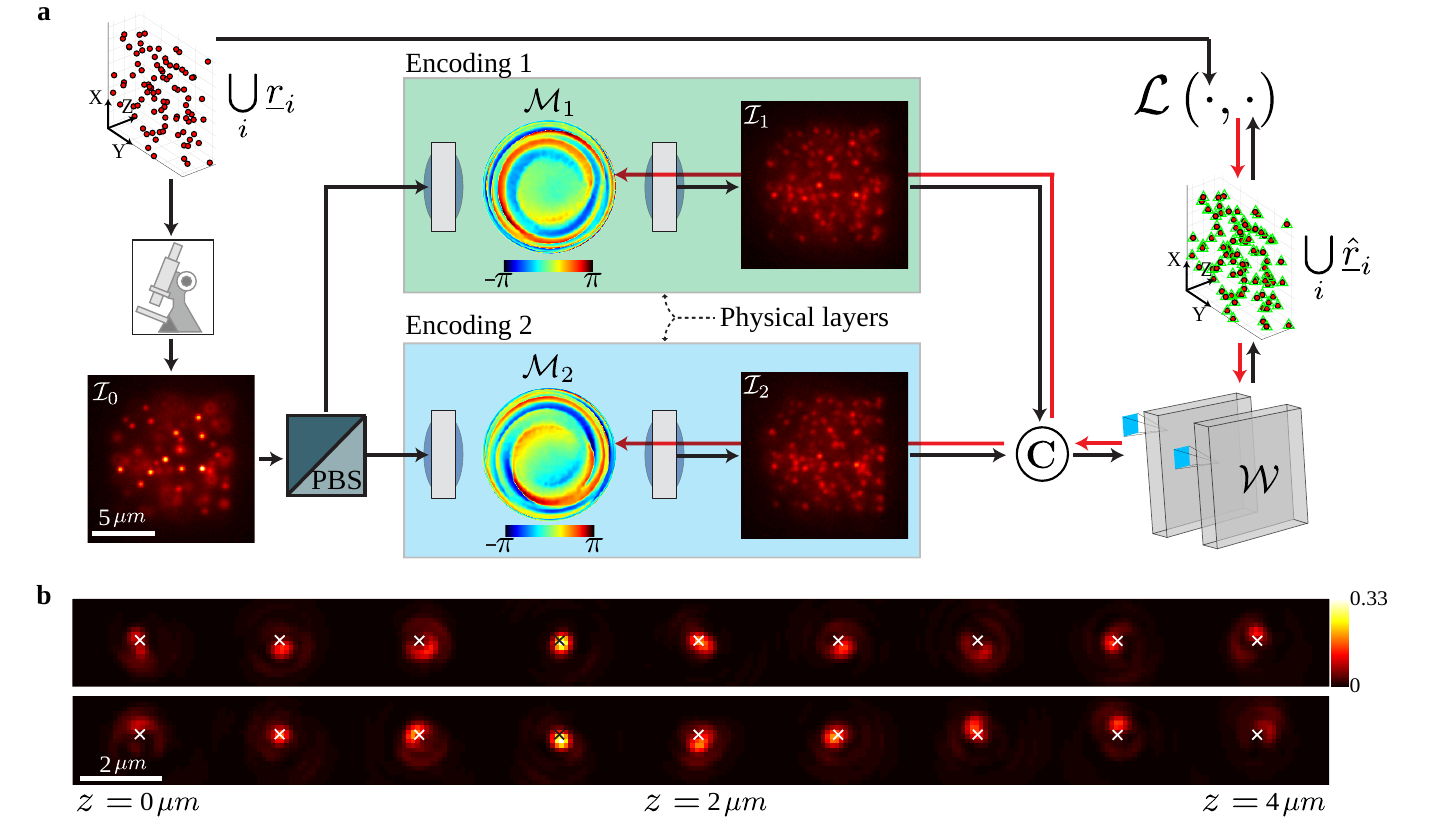}
\caption{End-to-end learning of the dual-channel optical system. (a) Simulated 3D positions of emitters \(\cup_i r_i\) are fed into two physical layers, which differ only in the applied phase mask \(\mathcal{M}\), to simulate the acquired image pairs with the modulated PSFs - \(\mathcal{I}_1\) \& \(\mathcal{I}_2\). Next, both images are fed through a convolutional neural network to recover the 3D positions in the simulation \(\cup_i \hat{r}_i\). Afterwards, these reconstructed positions are compared to the ground truth with our loss function \(\mathcal{L}\), and the gradients are back propagated through the layers (red lines) to jointly optimize the encoding masks \(\mathcal{M}_1\) \& \(\mathcal{M}_2\), and localization CNN parameters \(\mathcal{W}\). (b) Nebulae PSFs, which are the result of the end-to-end learning for a 4 \(\mu m\) axial range. The colorbar is normalized compared to the in-focus unmodulated PSF of the system.}
\label{fig:end-to-end-learning}
\end{figure*}

\subsection{End-to-end learning of a monocular PSF-pair} \label{subsec:end-to-end}

As shown previously \cite{nehme2020deepstorm3d}, end-to-end designs lead to efficient PSF patterns that are highly suited for dense 3D imaging. Here, we extend the DeepSTORM3D approach to tackle the problem of designing a PSF-pair. This is achieved by designing the encoding stage to incorporate two disjoint and differentiable physical-simulation layers (Fig. \ref{fig:end-to-end-learning}a). Each layer is parameterized by its own phase mask (\(\mathcal{M}_1\) \& \(\mathcal{M}_2\)) dictating the respective PSF (see supplementary section \ref{supp-imaging-model} for the imaging model). During training, we randomly simulate 3D positions (\(\cup_i r_i\)), and feed them to the two physical layers. Each physical layer encodes the 3D positions to their simulated sensor image (\(\mathcal{I}_1\) \& \(\mathcal{I}_2\)). These images are concatenated and fed to the localization CNN (parameterized by \(\mathcal{W}\)) which decodes them in order to recover the underlying 3D positions (\(\cup_i \hat{r}_i\)). The difference between the simulated and the recovered positions is quantified by our loss function (\(\mathcal{L}\)) and back-propagated to jointly optimize the phase masks (\(\mathcal{M}_1\) \& \(\mathcal{M}_2\)), and the localization CNN parameters (\(\mathcal{W}\)) end-to-end. This process is usually repeated for \(\approx\)30 epochs until convergence. For training details see supplementary section \ref{supp-learn-details}.

The end-to-end learned phase masks and their respective PSFs, hereafter referred to as the Nebulae PSFs, are presented in Fig. \ref{fig:end-to-end-learning}. Two distinctive features stand out in this pair compared to the previous approaches described earlier. First, both channels encode 3D information in their individual intensity patterns, as well as in the relative position of the intensity centroids throughout the entire axial range, a trait conceived to be useful for 3D localization before \cite{lew2011corkscrew}. Second, in phase-space, the learned phase masks are approximately rotated versions compared to one another, although our optimization was performed pixel-wise and our loss function did not include any constraints on the mutual information of both measurements.

To evaluate the performance of the Nebulae PSFs, we first compare them in simulation to the Tetrapod-EDOF pair (section \ref{sec:xy-z}), as well as to a single channel Tetrapod PSF with twice the signal (Fig. \ref{fig:nebulae-vs-tp-and-edof}). The results indicate that the Nebulae PSFs achieve unprecedented performance in localizing dense 3D emitters over a large axial range of 4 \(\mu m\)s assuming our experimental telomere imaging conditions, \emph{i.e}.\ \(\approx\)15K signal photons per emitter and \(\approx\)500 background photons per pixel.
\begin{figure*}[ht!]
\centering
\includegraphics{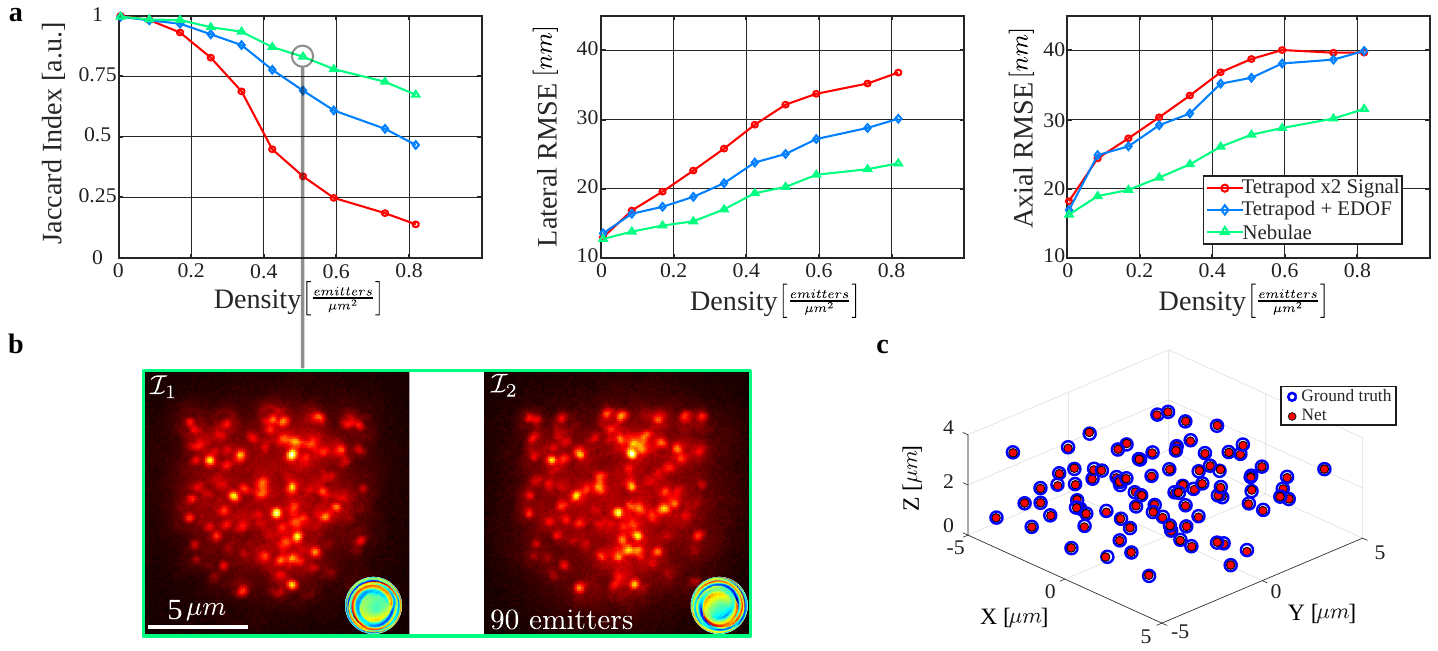} 
\caption{Performance as function of density. (a) Performance comparison of a single-channel Tetrapod (red), Tetrapod-EDOF pair (blue), and the Nebulae PSFs (green). The Nebulae PSFs performs best both in detectability (Jaccard index) and in precision (lateral/axial RMSE). Emitters were simulated with \(\approx\)15K signal photons per emitter and \(\approx\)500 background photons per pixel. Matching of points was computed with a threshold distance of 100 nm using the Hungarian algorithm. Each data point is an average of n = 100 simulated images. Average standard deviation in Jaccard index was \(\approx\)5\% and in precision was \(\approx\)3 nm. (b) Example of a simulated frame of density \(\approx\)0.5 \(\left[\frac{emitters}{\mu m^2}\right]\) alongside 3D comparison of the recovered (red) and the ground truth (blue) positions.}
\label{fig:nebulae-sim}
\end{figure*}

\subsection{Nebulae PSFs experimental validation} \label{subsec:nebulae-exp}

\begin{figure}[ht!]
\centering
\includegraphics[width=0.4\textwidth]{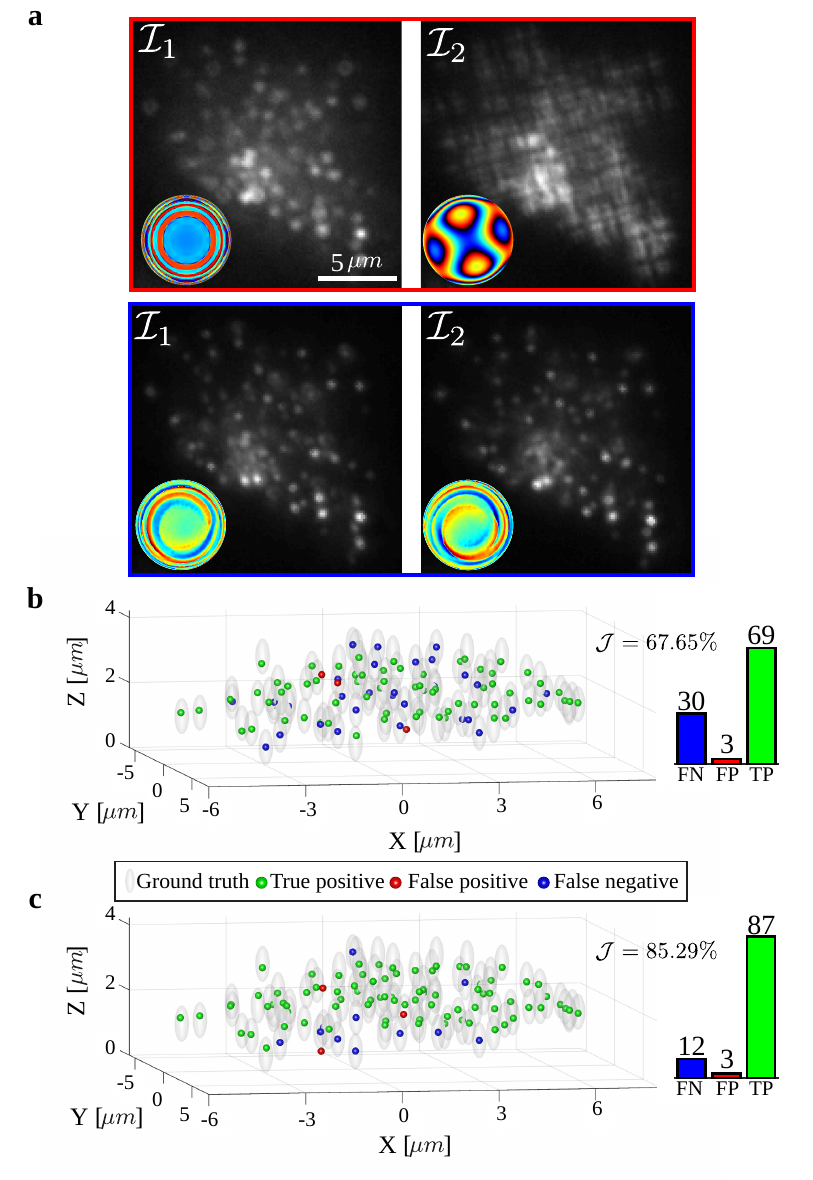} %
\caption{Experimental measurement of fixed U2OS cells with fluorescently labelled telomeres. Example images showing the two proposed mask pairs: the Tetrapod + EDOF (left) and the end-to-end learned pair (right). (b) The single-frame 3D localizations with the ground truth (achieved \textit{via} axial scanning) for the Tetrapod + EDOF and learned pair, respectively.} 
\label{fig:nebulae-vs-tp-and-edof}
\end{figure}

Next, we applied the Nebulae PSFs in fixed cells, and compared the performance to the Tetrapod-EDOF pair experimentally (Fig. \ref{fig:nebulae-vs-tp-and-edof}). Similar to section \ref{subsec:exp-tp-edof}, we first found the emitter positions by axial scanning, for comparison to our snapshots images taken at a single focal plane: once with the Tetrapod-EDOF pair, and once with the Nebulae PSFs. The results show that at a density of \(\approx\)0.34 \(\left[\frac{emitters}{\mu m^2}\right]\), the Nebulae PSFs are superior in localizing overlapping telomeres as measured by the Jaccard index. The Nebulae PSFs were also found to have superior performance relative to the CRLB-optimized pair from section \ref{subsec:crlb}. For a head-to-head comparison in simulations as well as experiments see supplementary sections \ref{supp-add-sim} and \ref{supp-add-exp}.

\section{Live telomere tracking} \label{sec:live-tracking}

Throughout this work we have imaged and localized 3D positions of telomeres in \textit{fixed} cells to facilitate quantitative comparisons of the proposed solutions. However, more pertinent is the application of our method to multiple-particle-tracking in \textit{live} cells, where axial scanning is inapplicable due to the motion of the objects. Here, our simultaneous multi-channel snapshot approach enables capturing the behavior of diffusing telomeres in living cells at an unprecedented combination of density, speed, and axial range \cite{weiss2018observing}.

Quantifying telomere dynamics in live cells is of paramount importance for answering fundamental questions under normal and disease conditions \cite{weiss2018observing,bronshtein2015loss}, as tracking the 3D diffusion of telomeres unveils information on the chromatin environment and on DNA folding regulation. One challenge in observing chromatin in living cells is the intrinsic biological heterogeneity between diffusing telomeres \cite{weiss2020three}. Therefore, to fully characterize chromatin dynamics it is desired to capture all single telomere trajectories, including in emitter-dense regions.

\begin{figure*}[ht!]
\centering
\includegraphics{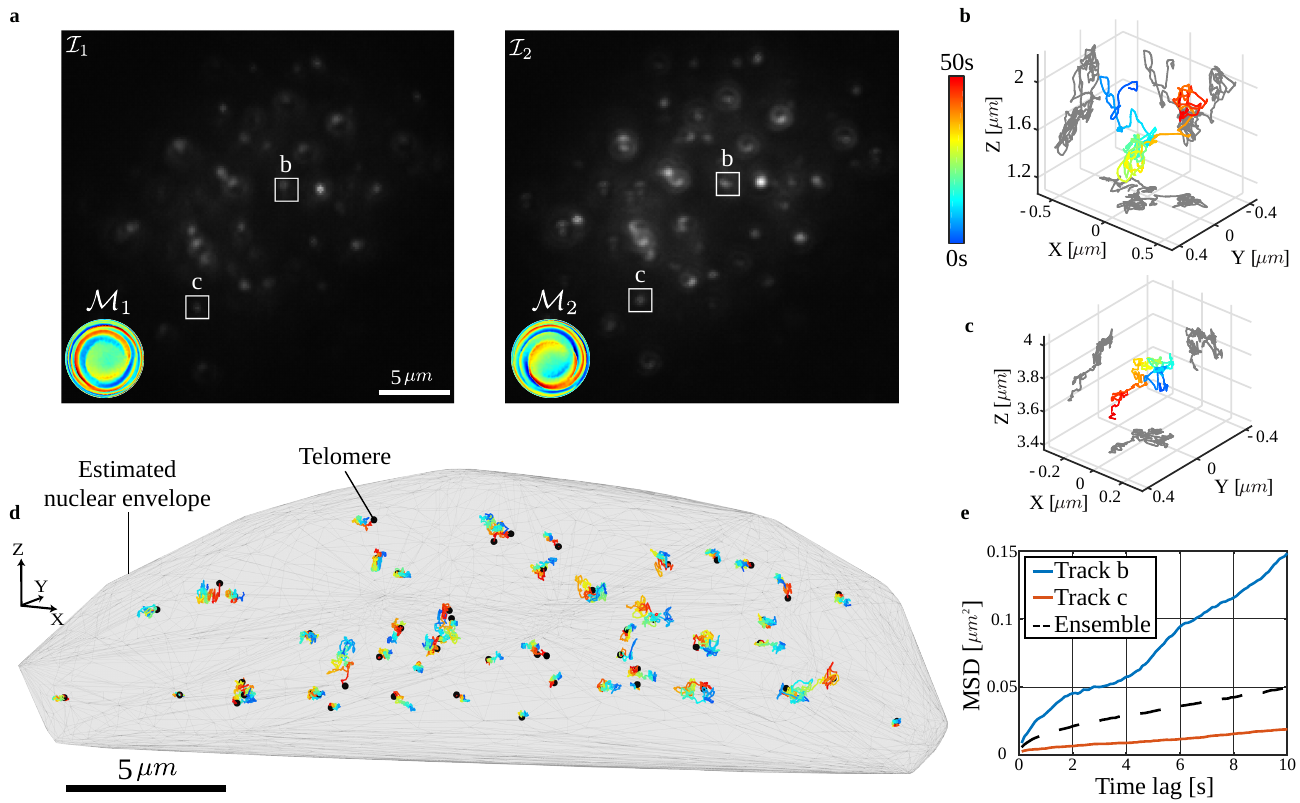}
\caption{Dense-particle tracking of labelled telomeres in live cancer cells with the Nebulae PSFs. (a) A single time point showing the two PSF-modulated images. (b)-(c) 3D spatiotemporal trajectories for telomeres (b) and (c), exhibiting drastically different diffusion behaviors, in different regions of the nucleus. (d) 3D rendered cell with all the accumulated tracks showing the motion tracking of telomeres in 3D. Most telomeres were localized in all frames (\(\leq 5\%\) missing localizations). (e) Ensemble MSD of all the estimated tracks, obscures the dynamics of individual particles, such as tracks (b) and (c), which exhibit very different diffusion dynamics.}
\label{fig:tracking}
\end{figure*}

Figure \ref{fig:tracking} demonstrates the full applicability of the Nebulae PSFs for volumetric tracking of \(\approx\)61 diffusing telomeres, spanning an axial range of \(\approx\)4.7 \(\mu m\) in the nucleus of a living U2OS cell. The trained localization CNN is able to reliably track all of the labelled telomeres over the course of 500 frames (50 s), even those in close proximity, and with a low signal-to-noise ratio. As evident in the resulting tracks (Fig. \ref{fig:tracking}), the telomeres exhibit variable diffusion profiles (Fig. \ref{fig:tracking}e) necessitating individual processing as facilitated by our approach.

\section{Discussion}

In computational imaging, the \emph{co-design} of optics and image-processing algorithms has been introduced in various applications spanning the fields of computational photography and computational microscopy. In the realm of localization microscopy, this is the key concept in PSF engineering \cite{huang2008three,pavani2009three,shechtman2014optimal}, and has been utilized to extend the imaging capabilities in SMLM \cite{aristov2018zola,shechtman2016multicolour}. Until recently, however, the standard approach was to design the optical system to optimize a specific trait of the PSF that would facilitate its processing afterwards, \emph{e.g}.\ an axial-displacement-induced rotation in the Double-Helix PSF \cite{schechner1996wave,pavani2009three}. In addition to conceived physical properties, information-content-driven optimization was also used in PSF-design; for example, in \cite{levin2007image}, where the PSF was optimized for depth discrimination. Similarly, for SMLM applications \cite{shechtman2014optimal} the PSF has been optimized to minimize the variance of an unbiased estimator for localizing the 3D position of a point source. While the latter two identified theoretically optimal solutions to encode the information, in complex environments, the decoding step is often limiting the problem as well.

Recently, powered by deep learning and differentiable physical models, \emph{end-to-end} designs of physical elements and data-processing algorithms have been demonstrated by our group and others to facilitate efficient imaging modalities in microscopy \cite{horstmeyer2017convolutional,muthumbi2019learned,kellman2019data,pinkard2020learned}. Specifically in SMLM,  the efficiency of jointly designing PSFs and deep networks was demonstrated for multi-color 2D imaging \cite{hershko2019multicolor} and snapshot dense 3D imaging \cite{nehme2020deepstorm3d}.

In this work, we addressed the challenging task of multi-PSF engineering for dense 3D imaging. Specifically, we proposed three different PSF-pairs, each derived with a different set of considerations. For the first pair, we introduced an efficient and laterally-compact EDOF PSF to complement the Tetrapod PSF at high emitter densities. Notably, this EDOF PSF has numerous applications in its own right for imaging in thick samples with little need for deconvolution \cite{dowski1995extended}. In the second pair, we extended the CRLB-design metric to optimize the sensitivity of a PSF-pair in the single-emitter case. Lastly, we presented the Nebulae PSFs, learned end-to-end to achieve reliable dense 3D localization \textit{via} from snapshot measurements. We validated each of the proposed designs numerically and experimentally. To demonstrate the applicability for dense 3D tracking in live cells, we tracked regions of dense telomeres using the Nebulae PSFs, enabling a statistical analysis of population heterogeneity, and high-resolution 3D modelling of chromatin dynamics in single cells.

In contrast to standard CNN filters, a notable aspect of end-to-end learning with physical layers, is our ability to visualize and interpret the designed physical elements. For example, for the Nebulae PSFs, the signal photons are compacted into a single lobe in each channel. This feature is understandably advantageous in the dense fields of emitters with limited SNR used in our simulations and experimental conditions. Moreover, the intensity patterns at each axial position combine elementary depth-encoding aberrations, such as astigmatism, rotation, and relative inter-channel single lobe movement. What separates these PSFs from predetermined designs is the simultaneous deployment of multiple depth-encoding strategies making full use of the decoding CNN capacity, and thereby optimizing dense 3D localization from noisy measurements.

Notably, our approach is not limited to particle tracking. By tweaking the physical-simulation layers, this method can be readily adapted to any point-source-sensing paradigm, including DAISY \cite{cabriel2019combining}, MINFLUX \cite{balzarotti2017nanometer,gwosch2020minflux}, multi-plane microscopy \cite{juette2008three,ram2008high,amin2020localization,louis2020fast}, and more. In a concurrent work \cite{ikoma2020snapshot}, similar ideas were pursued for multiplane PSF engineering demonstrating promising results in simulations. In these, and for SMLM applications, it is likely that modifying the CNN architecture, initializations, training sets, and loss functions, may further improve the performance, raising questions of how globally optimal is the solution derived in our framework. At this point, it is unclear how each optimization component affects the learning process, a question that will be addressed in future work. In particular, we anticipate that the emerging suite of tools developed to make deep learning more accessible to the community will assist in answering these critical questions \cite{ouyang2019imjoy,gomez2019deepimagej,von2020zerocostdl4mic}. 

To the best of our knowledge, this work reports the first end-to-end learning of multiple PSFs with experimental feasibility. Such multi-PSF designs may prove useful outside the realm of computational microscopy. For example, in computational photography, the design of coded aperture pairs and their optimal combination with stereo imaging has been a long standing question \cite{zhou2009coded,zhou2009good,zhou2011coded,levin2010analyzing,takeda2013fusing,gil2019monster}. Most recently, Gil \emph{et al}.\ \cite{gil2019monster} proposed to exploit identical phase-mask pairs for improved depth estimation and online stereo calibration. We believe this work paves the way for asymmetric strategies in the field of computational photography, with applications in stereo imaging, and multi-shot monocular depth estimation. Depending on the specific task at hand, the optimal PSF-pair could vary, however, we believe that the approaches to PSF-pair optimization in this work will provide a useful initialization to the general problem.

\clearpage
\section*{Funding}
The Israel Science Foundation (grant 852/17), the Technion Ollendorff Minerva Center, the Zuckerman Foundation. H2020 European Research Council Horizon 2020 (802567), Google Faculty Research Award for Machine Perception, The Israel Science Foundation (grant 450/18).

\section*{Acknowledgments}
We thank Rotem Mulayoff for insights and fruitful discussions with respect to the EDOF design. We also thank Romain F. Laine for his help with conceiving the name Nebulae. We gratefully acknowledge the support of the NVIDIA Corporation with the donation of the Titan Xp and the Titan V GPU used for this research. We thank Google for the cloud units provided to accelerate this research.

\section*{Disclosures}
The authors declare no conflicts of interest.


\appendix
\section{Appendix}

\renewcommand{\thefigure}{S\arabic{figure}}
\renewcommand{\theequation}{S\arabic{equation}}
\setcounter{figure}{0}
\setcounter{equation}{0}

\subsection{Imaging model} \label{supp-imaging-model}
In this section we briefly review the imaging model used throughout this work. Our system is composed of fluorescent emitters with an emission wavelength \(\lambda\) suspended in water (refractive index of \(n_2\approx 1.33\)) above an oil-immersed objective (refractive index of \(n_1\approx 1.518\)). The emitters are imaged with an objective lens (numerical aperture of NA), focused at a focus plane \(f\), and their image is magnified onto the sensor with a microscope magnification M. Let \(\mathcal{M}\) denote the phase mask placed in the conjugate back focal plane of an extended emission path with a 4\(f\) system (Fig. \ref{fig:optical-stup}), and let \(\left(\rho,\phi\right)\) denote the normalized radial coordinates in the Fourier plane such that \(\rho=1\) at \(\frac{\text{NA}}{n_1}\). Under the scalar approximation \cite{goodman2005introduction}, the PSF of a point source located at \(\left(x_0,y_0,z_0\right)\) above a water-oil interface is given by
\begin{equation}
    \text{PSF}^{th}\left(x,y;\mathcal{M},x_0,y_0,z_0\right) \propto \left|\mathcal{F}\left(\mathcal{A}\left(\rho,\phi\right)e^{j\mathcal{M} + \frac{2\pi j}{\lambda}\Phi\left(x_0,y_0,z_0,f\right)}\right)\right|^2,
    \label{eq:psf}
\end{equation}
where \(\left(x,y\right)\) are the coordinates at the image plane, \(\mathcal{F}\) is the two-dimensional Fourier transform, \(\mathcal{A}\left(\rho,\phi\right)\) is the effective aperture of the compound system, limited by \(n_2\) for high NA objectives
\begin{equation}
    \mathcal{A}\left(\rho,\phi\right) =
    \begin{cases}
    1,& \text{if } \rho\leq \frac{n_2}{n_1}\\
    0,& \text{otherwise}
    \end{cases},
    \label{eq:circ}
\end{equation}
and \(\Phi\left(x_0,y_0,z_0,f\right)\) is the accumulated phase due to the emitter 3D position and the focal plane setting. This phase can be decomposed into lateral and axial components
\begin{equation}
    \Phi\left(x_0,y_0,z_0,f\right) = \Phi_{xy}\left(x_0,y_0\right) + \Phi_{z}\left(z_0,f\right).
    \label{eq:phase3d}
\end{equation}
The lateral component is assumed to be a linear phase (\emph{i.e}.\ shift-invariant convolution system), given by
\begin{equation}
    \Phi_{xy}\left(x_0,y_0\right) = \frac{\text{M}\cdot\text{NA}}{\sqrt{\text{M}^2 - \text{NA}^2}}\left(x_0 \rho \cos{\phi} + y_0 \rho \sin{\phi}\right)
    \label{eq:phasexy}.
\end{equation}
As for the axial component, it is split into two terms to account for refractive index-mismatch \cite{hell1993aberrations}: the phase accumulated in water due to the emitter depth \(z_0\), and the phase accumulated in oil due to a focus shift \(f\) from the coverslip
\begin{equation}
    \Phi_z\left(z_0,f\right) = \Phi_{water}\left(z_0\right) + \Phi_{oil}\left(f\right),
    \label{eq:phasez}
\end{equation}
where,
\begin{align}
\begin{split}
    \Phi_{water}\left(z_0\right) &= z_0 n_2 \sqrt{1 - \left(\frac{n_1}{n_2}\right)\rho^2},\\
    \Phi_{oil}\left(f\right) &= - f n_1 \sqrt{1 - \rho^2}.
    \label{eq:phasewateroil}
\end{split}
\end{align}
Finally, the PSF in \cref{eq:psf} is slightly smoothed in image space
\begin{equation}
    \text{PSF}\left(x,y;\mathcal{M},x_0,y_0,z_0\right) = \text{PSF}^{th}\left(x,y;\mathcal{M},x_0,y_0,z_0\right) \circledast \mathcal{G} \left(x,y\right),
    \label{eq:psf-blur}
\end{equation}
Where \(\circledast\) denote convolution, and \(\mathcal{G}\left(x,y\right)\) is a 2D Gaussian kernel, with a standard deviation that is fit empirically to match experimental data (usually \(\approx\) 70 nm). This blur accounts for the finite size of the emitter, its spectrum, and the inherent blur in the optical system, alleviating the need to explicitly model these effects.  
For a full derivation of the model that includes neglected dipole and near-field effects, the reader is referred to \cite{axelrod2012fluorescence, backer2014extending}.

The image \(V\left(x,y\right)\) of a set of emitters \(\cup_{i} r_i\) is given by the incoherent sum of their PSFs
\begin{equation}
    V\left(x,y;\mathcal{M},\cup_i r_i\right) = \sum_i {\text{PSF}\left(x,y;\mathcal{M},r_i\right)}
    \label{eq:psf-sum},
\end{equation}
where \(r_i=\left(x_i,y_i,z_i\right)\) is the 3D position of the \(i^{\text{th}}\) emitter. 

The commonly used measurement model is given by a data-dependant Poisson noise, and an additive Gaussian read noise
\begin{equation}
    \mathcal{I}\left(x,y\right) \sim \mathcal{P}\left(V\left(x,y\right) + \mathcal{B}\left(x,y\right)\right) + \mathcal{N}\left(\mu,\sigma^2\right),
    \label{eq:measure-model}
\end{equation}
where \(\mathcal{P}\) is the Poisson distribution, \(\mathcal{B}\left(x,y\right)\) is a per-pixel background noise, \(\mathcal{N}\) is the the normal distribution, \(\mu\) is a baseline count level, and \(\sigma^2\) is the read-noise variance.

To make the measurement model differentiable, by the law of large numbers, we can approximate the Poisson noise with a Gaussian noise using the central limit theorem 
\begin{align}
\begin{split}
    & \mathcal{P}\left(V\left(x,y\right) + \mathcal{B}\left(x,y\right)\right) \approx \\ & \approx \mathcal{N}\left(V\left(x,y\right) + \mathcal{B}\left(x,y\right), V\left(x,y\right) + \mathcal{B}\left(x,y\right)\right).
    \label{eq:noise-approx}
\end{split}
\end{align}
The resulting data-dependant noise approximation is implemented using the reparameterization trick \cite{kingma2013auto}
\begin{align}
\begin{split}
    & \mathcal{N}\left(V\left(x,y\right) + \mathcal{B}\left(x,y\right), V\left(x,y\right) + \mathcal{B}\left(x,y\right)\right) \sim \\ & \sim V\left(x,y\right) + \mathcal{B}\left(x,y\right) + \sqrt{V\left(x,y\right) + \mathcal{B}\left(x,y\right)} \times \epsilon,
    \label{eq:noise-reparam}
\end{split}
\end{align}
where \(\epsilon\) is a realization of a standard normal distribution
\begin{equation}
    \epsilon \sim \mathcal{N}\left(0,1\right).
    \label{eq:epsilon}
\end{equation}

Now, the measurement model is differentiable \emph{w.r.t}.\ the phase mask \(\mathcal{M}\) and is therefore suited for end-to-end learning.

\subsection{EDOF PSF design} \label{supp-edof-design}
In this section we provide the implementation details for designing the EDOF PSF, then compare our result with existing popular designs. There are several ways to implement an EDOF PSF, including: placing an axicon in the optical path \cite{dufour2006two}, using ring apertures, and reducing the numerical aperture of the system. Due to photon-efficiency considerations, in this work we focus on the implementation of an EDOF PSF using a phase mask. Our general strategy is to formulate the design problem as a phase-retrieval task as detailed next. 

First, we start by simulating the in-focus Airy disk PSF for the desired optical system. Afterwards, this PSF is thresholded to keep only the main lobe with diameter D, and the result is fitted with a 2D Gaussian \(\mathcal{A}\left(x,y\right)\). This Gaussian is then replicated to generate a synthetic z-stack with 200 nm jumps between slices. \(\mathcal{A}\left(x,y\right)\) is also used to define a weighting matrix \(\mathcal{S}\left(x,y\right)\), that "squeezes" signal photons quickly into the diffraction limited spot, \(D\). Let \(\left(x,y\right)\) be centered pixel coordinates in image space, the matrix \(\mathcal{S}\left(x,y\right)\) is given by
\begin{equation}
    \mathcal{S}\left(x,y\right) =
    \begin{cases}
    1,& \text{if } \sqrt{x^2 + y^2}\leq D\\
    \alpha \cdot \sqrt{x^2 + y^2},& \text{otherwise}
    \end{cases},
    \label{eq:Ws-explicit}
\end{equation}
where in our implementation \(\alpha=25\), and \(D=150\ \left[\text{nm}\right]\) determined empirically to achieve appealing results.

Given \(\mathcal{S}\left(x,y\right)\), we try to retrieve the corresponding phase mask associated with the synthetic z-stack \textit{via} phase retrieval \cite{ferdman2020vipr}. This is implemented using Stochastic Gradient Descent (SGD) with importance sampling to minimize the following cost function
\begin{equation}
    \mathcal{L}_{\text{EDOF}}\left(\mathcal{M}\right) = \sum\limits_{i=1}^{N_z}{\|\left(PSF\left(x,y;\mathcal{M},z_i\right) - \mathcal{A}\left(x,y\right)\right)\cdot \mathcal{S}\left(x,y\right)\|_2^2},
\label{eq:L-edof}
\end{equation}
where \(\text{PSF}\left(x,y;\mathcal{M},z_i\right)\) is the on-axis PSF at depth \(z_i\), and \(N_z\) is the number of axial slices (\(\frac{\Delta_z}{200 \text{nm}}\)). Let \(\mathcal{Z}\left(\mathcal{M}\right)\) denote the current PSF stack dictated by phase mask \(\mathcal{M}\), such that \(\mathcal{Z}_i\left(\mathcal{M}\right)\equiv \text{PSF}\left(x,y;\mathcal{M},z_i\right)\). Our optimization is comprised of the 3 following steps:
\begin{enumerate}
    \item We compute the correlation \(\mathcal{C}\left(z\right)\) of \(\mathcal{Z}\left(\mathcal{M}\right)\) with \(\mathcal{A}\left(x,y\right)\) at each axial slice \(z_i\) 
\begin{equation}
    \mathcal{C}\left(z_i\right) = \left<\mathcal{A}\left(x,y\right),\mathcal{Z}_i\left(\mathcal{M}\right)\right>
\label{eq:corr}
\end{equation}
and choose the three axial slices \(\left(z_1,z_2,z_3\right)\) with the lowest correlation.
\item To avoid overfitting the sampled 200 nm ``knots" throughout the axial range, we perturb each of \(\left(z_1,z_2,z_3\right)\) locally with a random continuous shift \(\delta_z \sim \mathcal{U}\left[-100\text{nm},100\text{nm}\right]\) while clipping out-of-range values.
\item We calculate the gradient of the cost in \cref{eq:L-edof} sampled only at \(\left(z_1,z_2,z_3\right)\), and take a gradient step.
\end{enumerate}
In the third step, we experimented with a few adaptive SGD optimizers \cite{kingma2014adam,xiao2014proximal,dozat2016incorporating,liu2019variance}, and ultimately chose Adam \cite{kingma2014adam}. The process is repeated for 400 iterations, or till the loss function stagnates.

Notably, the correlation in our implementation serves as side-information \cite{gopal2016adaptive}, and is used to adaptively sample \(z\) slices and direct the SGD iterations. Compared with a stochastic sampling approach, this has the benefit of accelerating convergence, and empirically led to better solutions.

\begin{figure}[h!]
\centering
\includegraphics{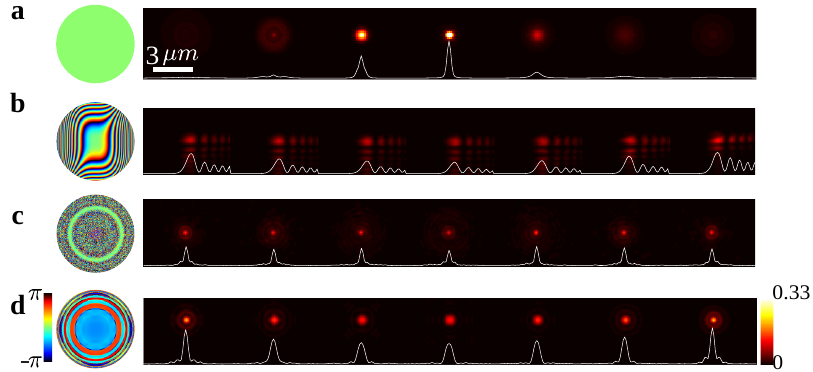}
\caption{Comparisons of EDOF PSFs in simulation. (a) Standard unmodulated PSF. (b) Cubic phase mask PSF. (c) Randomly sampled Fresnel lenses PSF. (d) Ours. The colorscale is normalized to the maximum intensity of the in-focus, unmodulated PSF.}
\label{figs:edof-comps}
\end{figure}

Figure \ref{figs:edof-comps} compares the result to two common EDOF implementations: the cubic phase mask \cite{dowski1995extended}, and the randomly sampled Fresnel lenses phase mask \cite{ben2005experimental}. The amplitude of the cubic phase mask was chosen such that the PSF is consistent over the FOV, but retains as much SNR as possible. Our proposed EDOF has three significant advantages over the classical designs: (1) its lateral extent is much smaller than the cubic phase mask PSF, matching our density requirements, (2) the SNR in the main spot is higher than both other methods, and (3) the proposed phase mask is smooth compared to randomly sampled Fresnel lenses. This facilitates its implementation using LC-SLM devices as these suffer from inter-pixel cross-talk \cite{moser2019model}.

\subsection{CRLB optimization} \label{supp-crlb-opt}
In this section we derive the  Cram\'er Rao Lower Bound (CRLB) \cite{kay1993fundamentals,ober2004localization,shechtman2014optimal} of the system in Fig. \ref{fig:optical-stup}. For simplicity, we start with the assumption that the measurement model is reduced to a Poisson data-dependent noise only. At the end, we also provide the expression for the extended case including the read noise. 

First let us start with some notation. Let \(\theta=\left(x_0,y_0,z_0\right)\) denote the 3D position of a single emitter imaged with the system in Fig. \ref{fig:optical-stup}, let \(u=\left(x,y\right)\) denote the concatenated coordinates at the image plane, and let \(\text{P}_{\theta}\left(u;\mathcal{M}_k\right)\equiv\text{PSF}\left(u;\mathcal{M}_k;\theta\right)\) denote the model PSF of the emitter in the detection path with phase modulation \(\mathcal{M}_k\). Assuming Poisson statistics for the source and background signals, the measured PSF \(\mathcal{I}_k\left(u\right)\) is given by
\begin{equation}
    \mathcal{I}_k\left(u\right) \sim \mathcal{P}\left(\text{P}_{\theta}\left(u;\mathcal{M}_k\right) + \mathcal{B}_k\left(u\right)\right),
\label{eq:poiss-only}
\end{equation}
where \(\mathcal{B}_k\left(u\right)\) is a per-pixel background. The log-likelihood function \(\ell\left(\mathcal{I}_k\left(u\right);\theta\right)\) for the measurement in \cref{eq:poiss-only} is given by
\begin{multline}
    \ell \left(\mathcal{I}_k\left(u\right);\theta\right) = \sum\limits_{u=1}^{\text{N}_u} \mathcal{I}_k\left(u\right)\cdot \log\left(\text{P}_{\theta}\left(u;\mathcal{M}_k\right)\right) - \text{P}_{\theta}\left(u;\mathcal{M}_k\right) \\
    + \text{C}\left(\mathcal{I}_k\left(u\right)\right),
\label{eq:loglik-poiss}
\end{multline}
where \(N_u\) is the number of pixels in the image, and \(\text{C}\left(\mathcal{I}_k\left(u\right)\right)\) is a function of the measurements that is independent of the unknown 3D position \(\theta\).

Given a log-likelihood function, the Fisher Information matrix \(\mathcal{F}\left(\theta\right)\) is defined as \cite{kay1993fundamentals}
\begin{equation}
\left[\mathcal{F}\left(\theta\right)\right]_{i,j} = \mathop{\mathbb{E}}\left[\left( \frac{\partial}{\partial \theta_i}\ell \left(\mathcal{I}_k\left(u\right);\theta\right)\right) \cdot \left( \frac{\partial}{\partial \theta_j}\ell \left(\mathcal{I}_k\left(u\right);\theta\right)\right)\Bigg{\lvert} \theta\right].
\label{eq:fisher-def}
\end{equation}
Substituting the log-likelihood from \cref{eq:loglik-poiss} we get
\begin{multline}
\left[\mathcal{F}\left(\theta;\mathcal{M}_k\right)\right]_{i,j} = \sum\limits_{u=1}^{\text{N}_u} \left(\frac{\partial}{\partial \theta_i}\text{P}_{\theta}\left(u;\mathcal{M}_k\right)\right) \cdot \left(\frac{\partial}{\partial \theta_j}\text{P}_{\theta}\left(u;\mathcal{M}_k\right)\right) \\ \cdot \frac{1}{\text{P}_{\theta}\left(u;\mathcal{M}_k\right) + \mathcal{B}\left(u\right)}.
\label{eq:fisher-poiss}
\end{multline}

Assuming independent photon arrivals in each detection path, the measurements \(\mathcal{I}_1\left(u\right)\),\(\mathcal{I}_2\left(u\right)\) become independent. Therefore, the joint information of both PSFs is given by the sum of the individual information from each PSF. Formally, let \(\mathcal{F}\left(\theta;\mathcal{M}_k\right)\) denote the information matrix of the measurement with phase modulation \(\mathcal{M}_k\). The joint Fisher Information matrix \(\mathcal{F}\left(\theta;\mathcal{M}_1,\mathcal{M}_2\right)\) for measurements \(\mathcal{I}_1\left(u\right)\),\(\mathcal{I}_2\left(u\right)\) is given by
\begin{equation}
    \mathcal{F}\left(\theta;\mathcal{M}_1,\mathcal{M}_2\right) = \mathcal{F}\left(\theta;\mathcal{M}_1\right) + \mathcal{F}\left(\theta;\mathcal{M}_2\right).
    \label{eq:F-sum}
\end{equation}

Let \(\theta_i \in \{x_0,y_0,z_0\}\) denote the coordinate of the 3D position. Given \(\mathcal{F}\left(\theta;\mathcal{M}_1,\mathcal{M}_2\right)\), the CRLB for estimating \(\theta_i\) is defined as \cite{kay1993fundamentals}
\begin{equation}
    \text{CRLB}_i\left(\theta;\mathcal{M}_1,\mathcal{M}_2\right) = \left[\mathcal{F}\left(\theta;\mathcal{M}_1,\mathcal{M}_2\right)^{-1}\right]_{i,i},
\label{eq:crlb}
\end{equation}
where \(\mathcal{F}\left(\theta;\mathcal{M}_1,\mathcal{M}_2\right)^{-1}\) denote the inverse of the Fisher information matrix. Based on \cref{eq:crlb}, to derive the phase masks \(\mathcal{M}_1\),\(\mathcal{M}_2\) optimizing the CRLB for all three estimated parameters \(\hat{\theta}=\left(\hat{x}_0,\hat{y}_0,\hat{z}_0\right)\), we minimize the following cost function
\begin{equation}
    \mathcal{L}_{\text{CRLB}}\left(\mathcal{M}_1,\mathcal{M}_2\right) = \sum\limits_{i\in\{\hat{x}_0,\hat{y}_0,\hat{z}_0\}}\sum\limits_{\theta'} \sqrt{\text{CRLB}_i \left(\theta';\mathcal{M}_1,\mathcal{M}_2\right)}.
\label{eq:Lcrlb}
\end{equation}

In our implementation, \(\text{CRLB}_i\left(\theta'\right)\) is evaluated at on-axis positions \(\theta'=\left(0,0,z'\right)\), where \(z'\) is sampled each 250 nm throughout the desired axial range. We also simplify the per-pixel background term \(\mathcal{B}\left(u\right)\) to a single scalar of 15 photons per pixel, and scale the PSFs to match realistic signal counts encountered in SMLM imaging, \emph{i.e}.\ 2000 photons per emitter. Notably, different from our previous work \cite{shechtman2014optimal}, we optimized the CRLB using a per-pixel approach rather than constraining the solution to a subspace of Zernike polynomials. This was particularly important to efficiently navigate the wide variety of possible solutions. 

Finally, in this work we focused our attention on SMLM experimental conditions. Therefore, for our purpose the read noise effects were negligible. However, the optimization is readily extended to the mixed Poisson-Gaussian case by revisiting \cref{eq:poiss-only,eq:loglik-poiss,eq:fisher-poiss}. Specifically, assume the measured PSF \(\mathcal{I}_k\left(u\right)\) is given by
\begin{equation}
    \mathcal{I}_k\left(u\right) \sim \mathcal{P}\left(\text{P}_{\theta}\left(u;\mathcal{M}_k\right) + \mathcal{B}_k\left(u\right)\right) + \mathcal{N}\left(\mu,\sigma^2\right),
\label{eq:poiss-gauss}
\end{equation}
where \(\mathcal{N}\) is the normal distribution, \(\mu\) is a baseline, and \(\sigma^2\) is the variance of the read noise. 

We can approximate the Poisson noise by a Gaussian noise using \cref{eq:noise-approx}
\begin{multline}
    \mathcal{I}_k\left(u\right) \sim \mathcal{N}\left(\text{P}_{\theta}\left(u;\mathcal{M}_k\right) + \mathcal{B}_k\left(u\right),\text{P}_{\theta}\left(u;\mathcal{M}_k\right) + \mathcal{B}_k\left(u\right)\right) \\ + \mathcal{N}\left(\mu,\sigma^2\right).
\label{eq:gauss-gauss}
\end{multline}
Assuming both noise sources are independent we get
\begin{equation}
    \mathcal{I}_k\left(u\right) \sim \mathcal{N}\left(\text{P}_{\theta}\left(u;\mathcal{M}_k\right) + \mathcal{B}_k\left(u\right)+\mu,\text{P}_{\theta}\left(u;\mathcal{M}_k\right) + \mathcal{B}_k\left(u\right)+\sigma^2\right).
\label{eq:one-gauss}
\end{equation}
The resulting log-likelihood function \(\ell\left(\mathcal{I}_k\left(u\right);\theta\right)\) for the measurement in \cref{eq:one-gauss} is given by
\begin{multline}
    \ell \left(\mathcal{I}_k\left(u\right);\theta\right) = - \sum\limits_{u=1}^{\text{N}_u} \log\left(\text{P}_{\theta}\left(u;\mathcal{M}_k\right) + \mathcal{B}_k\left(u\right) + \sigma^2\right) + \\ \frac{\left(\mathcal{I}_k\left(u\right) - \text{P}_{\theta}\left(u;\mathcal{M}_k\right)\right)^2}{\text{P}_{\theta}\left(u;\mathcal{M}_k\right) + \mathcal{B}_k\left(u\right) + \sigma^2},
\label{eq:loglik-gauss}
\end{multline}
where \(N_u\) is the number of pixels in the image. Substituting the log-likelihood from \cref{eq:loglik-gauss} in the definition from \cref{eq:fisher-def} we get
\begin{equation}
\begin{split}
& \left[\mathcal{F}\left(\theta;\mathcal{M}_k\right)\right]_{i,j} = \sum\limits_{u=1}^{\text{N}_u} \left(\frac{\partial}{\partial \theta_i}\text{P}_{\theta}\left(u;\mathcal{M}_k\right)\right)\cdot \left(\frac{\partial}{\partial \theta_j}\text{P}_{\theta}\left(u;\mathcal{M}_k\right)\right) \\ & \cdot \left(\frac{1}{\text{P}_{\theta}\left(u;\mathcal{M}_k\right) + \mathcal{B}\left(u\right) + \sigma^2} + \frac{1}{2\left(\text{P}_{\theta}\left(u;\mathcal{M}_k\right) + \mathcal{B}\left(u\right) + \sigma^2\right)^2}\right).
\label{eq:fisher-gauss}
\end{split}
\end{equation}

Substituting \cref{eq:fisher-gauss} in \cref{eq:F-sum,eq:crlb,eq:Lcrlb} we get the desired cost function for the general case.

\begin{figure*}[hbt!]
\centering
\includegraphics{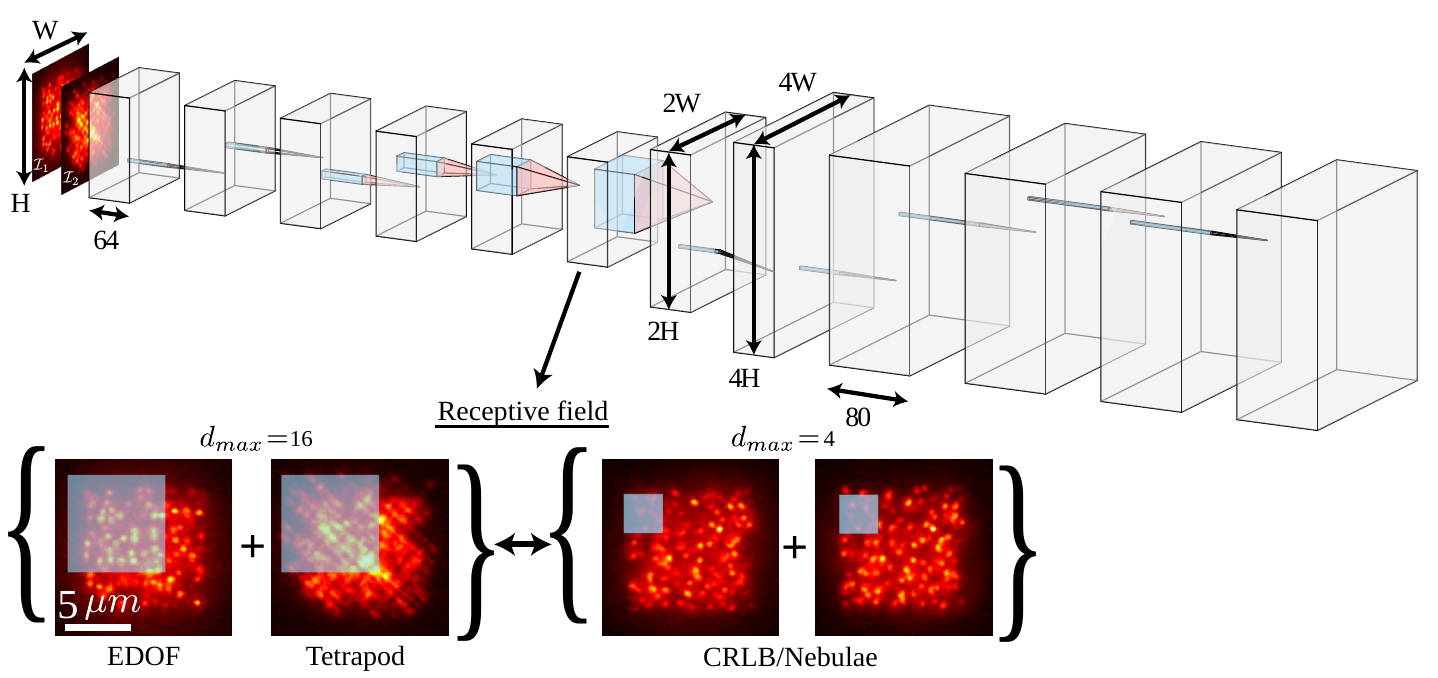}
\caption{CNN architecture. (a) The concatenated snapshot images \(\mathcal{I}_1\),\(\mathcal{I}_2\) are fed to a CNN composed of 3 modules as described in the text. Feature maps dimensions are depicted with \cite{lenail2019nn} to reflect the operation of each module. The spatial supports of all convolutional filters are \(3\times3\). The number of channels is fixed to 64 in both the multi-scale context aggregation, and the upsampling modules. Then, the number is increased to 80 for the refinement module. Note that in the context-aggregation module the spatial support of all convolutional filters is \(3\times3\), although their receptive field grows exponentially with the dilation rate. Blue square depicts the final receptive field for both choices of \(d_{max}\). The output 3D high-resolution volume is translated to a list of 3D localizations through simple post-processing. Scale bars are 3 \(\mu\)m.}
\label{figs:cnn-loc}
\end{figure*}

\subsection{Learning details} \label{supp-learn-details}
\subsubsection{CNN architecture}

In this work, we adapt the CNN architecture previously proposed in DeepSOTRM3D \cite{nehme2020deepstorm3d} to process an image with 2 channels (Fig. \ref{figs:cnn-loc}). Our architecture is relatively light with only \(\approx\)440K trainable parameters, comprised of 3 main modules:

\begin{enumerate}
\item Multi-scale context-aggregation module: we used dilated convolutions \cite{yu2015multi} to increase the receptive field of each layer while keeping a fixed number of 64 channels. The two concatenated snapshots are processed through 6 convolutional blocks with increasing dilation rates. The maximal dilation rate \(d_{max}\) was set according to the PSFs lateral footprint: \(d_{max}=16\) for the Tetrapod-EDOF pair, and \(d_{max}=4\) for the other two PSF pairs (see Fig. \ref{figs:cnn-loc}). We also include skip connections to improve gradient flow \cite{he2016deep} (not shown in the figure).

\item Upsampling module: composed of two consecutive \(\times 2\) resize-convolutions \cite{odena2016deconvolution} to increase the lateral resolution by a factor of 4. We used nearest-neighbor interpolation to resize the images. Assuming a CCD pixel-size of 110 nm, the lateral pixel-size of the upsampled features is 27.5 nm.

\item Prediction module: after super-resolving emitters in the lateral dimension, we further refine their axial position through 3 additional convolutional blocks with an increased number of channels. For a 4 \(\mu\)m range, we use 80 channels, \emph{i.e}.\ a voxel-size of \(\approx\)50 nm in \(z\). The final prediction is given by a \(1 \times 1\) convolution followed by an element-wise HardTanh to limit the output range to \(\left[0, \text{W}\right]\), where W is set empirically to 800 to account for class imbalance (occupied vs. vacant voxels).
\end{enumerate}

The spatial supports of all convolutional filters are \(3\times3\). Each convolution block is follow by a Batch Normalization layer, and a LeakyReLU non-linearity with slope \(\alpha=0.2\). Note that depth is exchanged with channels as our architecture is composed of solely 2D convolutional layers. Afterwards, these dimensions are permuted in the recovered volume. To compile a list of localizations at test time, we threshold the voxel values and find local maxima in clustered components (details in section \ref{supp-subsec-post-process}). Lastly, to efficiently learn the phase masks with reduced computation, we modify the architecture in a similar fashion to that described in \cite{nehme2020deepstorm3d}.

Notably, in this work we used the same encoder to process both images. In our implementation the image pair is first warped using a calibrated affine transform prior to CNN processing.
However, in case of severe inter-channel misalingment this is expected to be sub-optimal, and a ``Y-net" structure with separate encoders should be considered. In particular, one of the encoders could be potentially swapped with a spatial transformer network \cite{jaderberg2015spatial} to alleviate the need for calibration.

\subsubsection{Training set}

To learn a localization CNN solely with predefined phase masks, we simulate a training set composed of 10K simulated image-pairs and their corresponding labels which are lists of emitter positions. 9K examples were used for training with 1K examples held out for validation. Alternatively, to jointly learn the phase masks and the localization CNN parameters, the training set is composed of solely simulated emitter positions, as the respective image-pairs are being changed throughout iterations according to the phase masks.

In our implementation the training positions are randomly drawn within the 3D cube of possible locations in order for the method to generalize to arbitrary imaged structures. The Boolean grid used as label in training is given by projecting the continuous positions on the recovery grid (voxel size of \(27.5 \times 27.5 \times 50 \ nm^3\)).

Given a set of 3D locations, the expected model images are simulated using the measurement model in \cref{eq:measure-model}. To accurately model experimental data in our simulations, we image beads on the coverslip prior to the experiment, and retrieve the aberrated pupil functions using VIPR \cite{ferdman2020vipr}. To make our simulations realistic, we diversify the training conditions to include experimentally variability. Namely, we vary the emitter density, the signal-to-noise ratio, the amount of blur, and any additional expected experimental challenges (\emph{e.g}.\ motion blur, laser fringes etc.). For example, in telomere imaging we have observed a highly non-uniform per-pixel background, presumably resulting from the nucleus auto-fluorescence. To model this effect, we approximate the per-pixel background \(\mathcal{B}\left(x,y\right)\) in \cref{eq:measure-model} using a super-Gaussian
\begin{equation}
\mathcal{B}\left(u\right) = \mathcal{A}_1 e^{-\frac{1}{2}\left(u - \mu\right)^T \Sigma^{-1} \left(u-\mu\right)} + \mathcal{A}_2,
\label{eq:super-gauss}
\end{equation}
where \(u=\left(x,y\right)\) is the combined 2D coordinates in image space, \(\mathcal{A}_1\),\(\mathcal{A}_2\) are scaling parameters, \(\mu\) is the 2D centroid, and \(\Sigma\) is the covariance matrix. These parameters are augmented in training to make the model robust to their variations.

\begin{figure*}[ht!]
\centering
\includegraphics{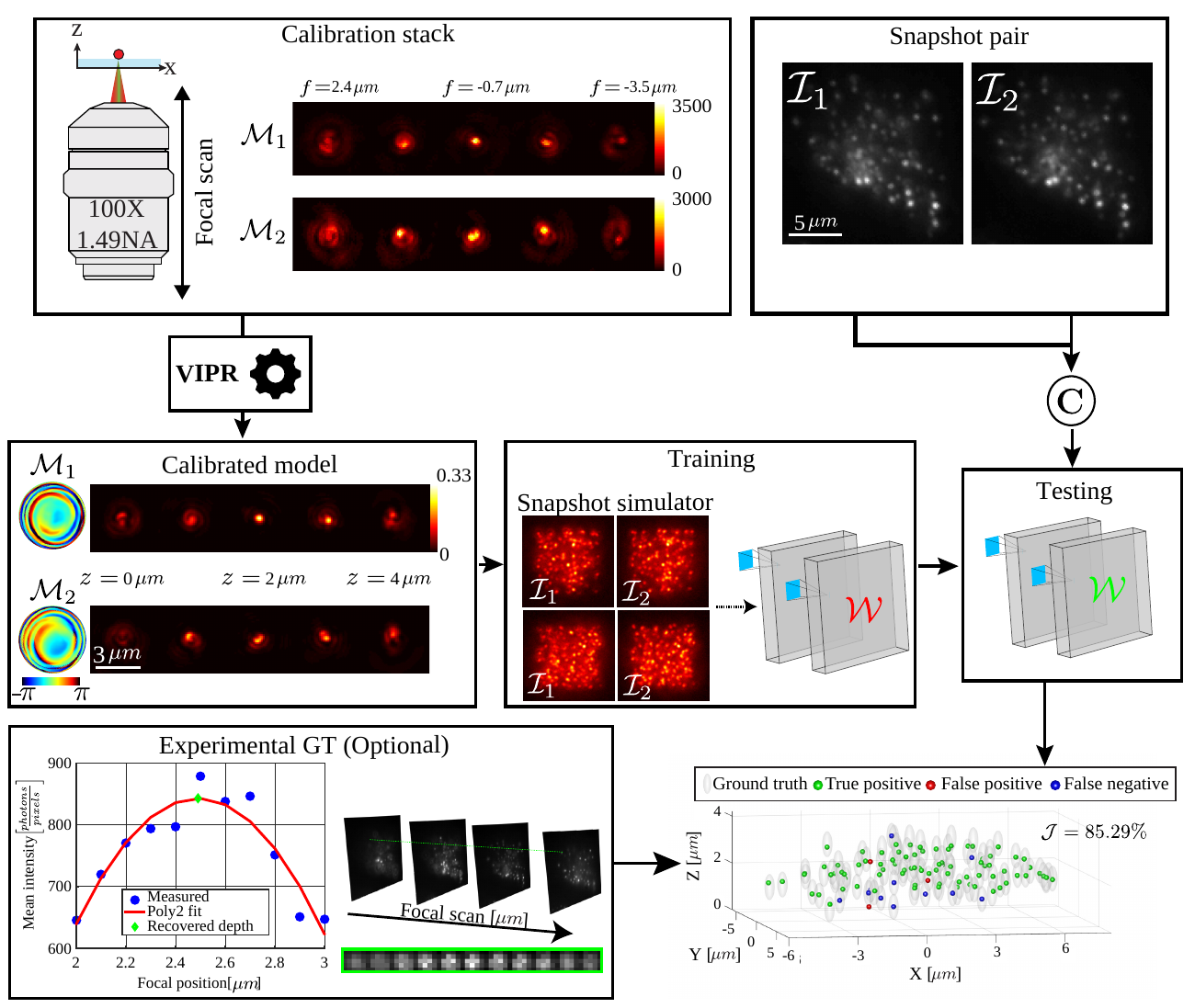}
\caption{Overview of a typical experiment. Fluorescent beads are used to create 3D PSF scans of the two channels, which are then modelled using VIPR. The calibrated PSF models are used to train the localization net. The trained net can then localize experimental data, and output the desired 3D positions from snapshot measurements. For fixed samples, where an experimental ground truth is available, the Jaccard index is calculated by matching the axial scan results with the net output.} 
\label{figs:calib-proc}
\end{figure*}

\subsubsection{Loss function}
Let \(x\) denote the GT boolean volume, and \(\hat{x}\) denote the network prediction. Our loss function for training the net \(\mathcal{L}_{Net}\) is a combination of two terms
\begin{equation}
\mathcal{L}_{Net}\left(x,\hat{x}\right) = \mathcal{L}_{Heatmaps}\left(x,\hat{x}\right) + \mathcal{L}_{Overlap}\left(x,\hat{x}\right).
\label{eq:loss-net}
\end{equation}
The first term \(\mathcal{L}_{Heatmaps}\) is a 3D heatmap matching loss, given by
\begin{equation}
\mathcal{L}_{Heatmaps}\left(x,\hat{x}\right) = \|x\circledast\mathcal{G}_{3D} - \hat{x}\circledast\mathcal{G}_{3D}\|^2,
\label{eq:loss-g3d}
\end{equation}
where \(\mathcal{G}_{3D}\) is a 3D Gaussian kernel with a standard deviation of 1 voxel. This term measures the proximity of our prediction to the simulated ground truth by measuring the \(\ell_2\) distance between their respective heatmaps. 

The second term \(\mathcal{L}_{Overlap}\) is a measure of overlap, given by
\begin{equation}
\mathcal{L}_{Overlap}\left(x,\hat{x}\right) = 1 - \frac{2\cdot\sum_i{x_i\cdot\hat{x}_i}}{\sum_i{x_i\cdot\hat{x}_i} + \sum_i{x_i}}. 
\label{eq:loss-overlap}
\end{equation}
This term provides a soft approximation of the true positive rate in the prediction. Note that \(\mathcal{L}_{Overlap}\) doesn't take into account false positives, and hence if optimized alone will result in a predicted volume of 1s. Although, here this is not a feasible solution as it is not favored by \(\mathcal{L}_{Heatmaps}\). In our implementation we weight voxels containing emitters with a factor of W=800 in order to balance out the contributions of vacant and occupied voxels. Hence, the CNN output is constrained to be in the range \(\left[0, 800\right]\). This strategy makes optimization easier and prevents gradient clipping.

\subsubsection{Optimization and hyper-parameters}

We used the Adam optimizer \cite{kingma2014adam} with the following parameters: \(\text{lr} = 5\times10^{-3}, \beta_1 = 0.9, \beta_2 = 0.999, \epsilon = 10^{-8}\). The batch size was 16 for learning a phase mask, and 4 for learning a recovery net (due to GPU memory). The learning rate was reduced by a factor of 10 when the loss plateaus for more than 5 epochs, and training was stopped if no improvement was observed for more than 7 epochs, or alternatively a maximum number of 50 epochs was reached. The initial weights were sampled from a uniform distribution on the interval \(\left[-\sqrt{k},\sqrt{k}\right]\) where \(k=\frac{1}{k_x \times k_y \times C_{in}}\), with \(k_x,k_y\) the filter spatial dimensions, and \(C_{in}\) the number of input channels to the convolutional layer. Training and evaluation were run on a workstation equipped with 32 GB of memory, an Intel(R) Core(TM) \(i7 - 8700\), 3.20 GHz CPU, and a NVidia GeForce Titan Xp GPU with 12 GB of video memory. Phase mask learning took \(\approx 25\) h, and recovery net training took \(\approx 35\) h. Our code is implemented using the Pytorch framework \cite{paszke2017automatic}, and soon will be made publicly available at \url{https://github.com/EliasNehme/DeepNebulae}.

\subsubsection{Post-processing} \label{supp-subsec-post-process}

The fully convolutional architecture that we adopted in this work outputs a super-resolved 3D volume, where occupied voxels account for emitters. To compile a list of localizations, we first threshold this volume keeping only voxels with a minimal confidence of 80 (maximal output is 800). Afterwards, out of the remaining localizations we discard those which are not local maximas in their 3D vicinity. The radius used for grouping and local maxima finding was \(\approx\)100 nm. Lastly, the recovered continuous 3D position is given by applying the 3D Center of Gravity (CoG) estimator to the vicinity of the local maximas in the prediction volume. While it is possible to use more sophisticated post-processing steps we choose to use this simple and efficient strategy to keep our method as fast as possible. In our implementation we write these steps as a composition of pooling and convolution operations, making calculations extremely efficient on GPU. 

Notably, While grouping and local maxima finding potentially limits the maximal density, keep in mind that overlaps in 2D normally translates to non-overlapping "blobs" in 3D. Hence, this is hardly a limitation in common imaging conditions as localization algorithms struggle considerably before reaching this limit.

In the telomere tracking experiment, the per-frame localizations were linked using DBSCAN clustering \cite{ester1996density} applied directly to the 3D positions. The maximum distance allowed between points was \(\epsilon=0.25 \mu m\), and the minimal number of emitters per cluster was minPts=25. This resulted in filtering 83 localizations out of 24530 throughout the 500 frames, \emph{i.e}.\ less than 0.3\%. All tracks started within the first 6 frames and were relatively clustered in 3D with no bifurcations observed. For more complicated tracking scenarios the reader is encouraged to link the CNN localizations by resorting to a more robust tracking software such as \cite{tinevez2017trackmate}.

\subsection{Experimental implementation} \label{supp-exp-implement}

This section details the full experimental procedure to localize emitters using snapshot measurements from the dual-view setup. An outline of a typical experiment is presented in Fig. \ref{figs:calib-proc}. The following subsections detail each part of the experiment for completeness.

\subsubsection{Dual channel calibration} \label{subsec:supp-calib-process}
The goal of this section is to describe the process of calibrating the proposed dual-camera system, such that simulated PSFs will match measured data and their positions will correspond between the two images. The practice of aligning an optical 4\(f\) Fourier processing system, calibrating a LC-SLM, and creating a simulated model for a single channel has been meticulously explained in many previous works (\emph{e.g}.\ \cite{siemons2018high}).

\begin{figure}[h!]
\centering
\includegraphics{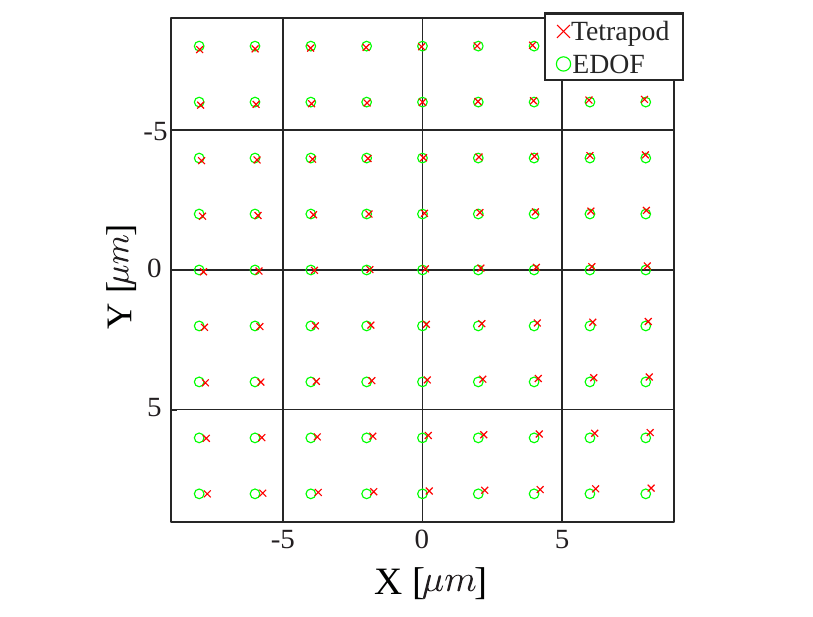}
\caption{Channel registration. The estimated affine transformation for the Tetrapod-EDOF experiment (main text Fig. \ref{fig:tp-vs-tp-edof}).}
\label{figs:affine-Trans}
\end{figure}

The proposed system consists of two identical optical paths which generate 3D PSF images. The acquired images are encoded simultaneously in the localization network, and thus pose some extra challenges in the calibration process, specifically with respect to their spatial alignment. In our work, post-processing corrections are not a viable option due to the density of PSFs, necessitating a good calibration of the 3D alignment. For this end, we created two calibration samples (sparse and dense) consisting of a water-covered glass coverslip (Fisher Scientific) with 40 nm fluorescent beads (FluoSpheres (580/605), ThermoFisher) adhered to the surface with 1\% PVA.
The dense sample was chosen such that the unmodulated PSFs will cover the entire field of view (FOV) but each individual bead can still be fit using ThunderSTORM \cite{ovesny2014thunderstorm}. The localizations from each channel were used to estimate an affine transformation between the two cameras (Fig. \ref{figs:affine-Trans}). To prevent outliers from biasing the transformation, we implemented a Random sample consensus (RANSAC) procedure.

Next, the sparse sample is chosen such that each slice of the 3D PSFs (for both channels) can be imaged without any overlaps from neighboring emitters. An axial scan is performed to ensure that both channels measure corresponding PSFs at the same focal plane positions, to account for any minor axial misalignment between the two cameras. The point of reference (lateral) was chosen as the center of gravity of the maximum projection in one of the channels.

The reference point of the second channel was calculated using the aforementioned affine transformation. Next, we used VIPR \cite{ferdman2020vipr} to generate a phase mask for each channel, as it provides with a good model and accounts for the issue of wobble and near field effects by implementing the vectorial diffraction model.
Importantly, while the affine transformation is calculated using localizations and not images, ultimately the input to the localization network is an image-pair. However, since a global affine transformation is not a shift-invariant operator, a fully convolutional model will struggle to learn this operator efficiently. Therefore, at test time, we warp the image of one camera to align with its counterpart, and feed the aligned concatenated image pair to the network. The warping operation is implemented using cubic-spline interpolation.

\begin{figure}[h!]
\centering
\includegraphics{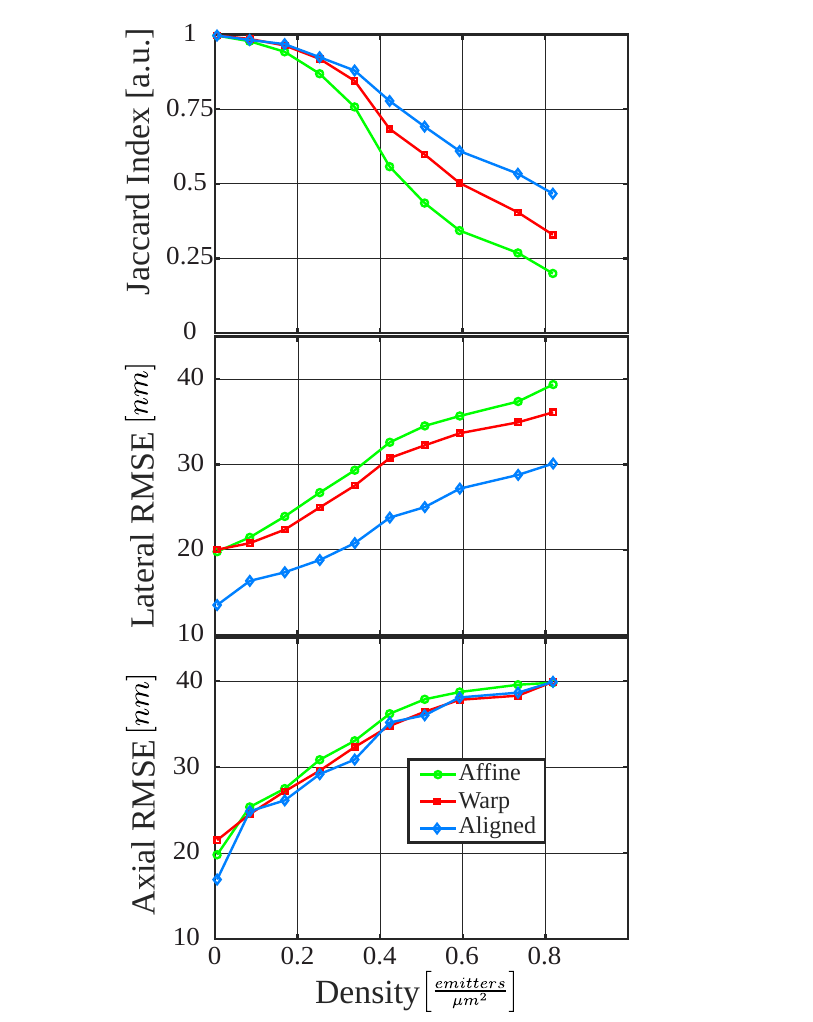}
\caption{Effect of image misalignment. Numerical comparison between networks trained with aligned images (blue), misaligned images (green) and approximately aligned images (up to 50 nm) by warping (red).}
\label{figs:image-misalign}
\end{figure}

To test the importance of image alignment, we trained three different models: (1) with perfectly aligned positions, (2) with randomly misaligned positions (achieve by sampling portions of the estimated transformation), and (3) with misaligned positions accompanied by a known transformation between channels (up to 50 nm) that is used to warp the images. Three conclusions can be made based on the results (Fig. \ref{figs:image-misalign}): (1) it is clear that the model is unable to efficiently cope with a random global transform, (2) calibrating the affine transform up to 50 nm errors and warping the images prior to localization improves performance, and (3) perfect alignment of the Tetrapod and the EDOF PSFs does not improve the axial localization precision. The latter is expected because the axial information is decoded solely based on the Tetrapod channel. Therefore, it is insensitive to the alignment with the EDOF PSF which does not encode \(z\).

\subsubsection{Optical components} \label{supp-subsec-optics-comp}

The imaging system in Fig. \ref{fig:optical-stup} consists of a Nikon Eclipse-Ti inverted fluorescence microscope with a 100X/1.49 NA Nikon objective (CFI SR HP Apo TIRF 100XC). A polarizing beam splitter was placed after the first achromatic doublet lens (f=15 cm) to split the emission path. Both paths consisted of three additional achromatic doubles lenses to image the back focal plane onto a LC-SLM (Pluto-VIS020, Holoeye in the first path, and 1920X1152 liquid crystal on silicon, Meadowlark in the second). After a last image-forming lens, the modulated images were recorded by two sCMOS cameras (Prime 95B, Photometrics). For full synchronization, the first camera triggered the second camera (in a leader-follower configuration), which in turn triggered the 561 nm illumination laser (iChrome MLE, Toptica). 

\subsubsection{Biological sample preparation}

For cell experiments, U2OS cells were prepared as described previously in \cite{nehme2020deepstorm3d}. In brief, cells were grown in standard conditions: \(37^o C\), 5\% \(\text{CO}_2\) in Dulbecco`s Modified Eagle Media (DMEM - without phenol red for the live cells experiment) with 1 \(\text{gl}^{-1}\) D-glucose (low glucose), supplemented with 10\% fetal bovine serum, and 1\% penicillin–streptomycin and glutamine. To fluorescently label the telomeres, cells were transfected with a plasmid encoding the fluorescently tagged telomeric repeat binding factor 1 (DsRed-hTRF1) using Lipofectamine 3000 (Thermo Fisher Scientific). After 20-24 hours, cells were either fixed with 4\% paraformaldehyde for 20 min, washed three times with PBS and mounted to a slide (\(22\times22\) \(\text{mm}^2\), 170 \(\mu m\) thick) with mounting medium; or imaged live in a temperature, humidity, and gas-mixture controlled imaging chamber mounted to the microscope (Okolab) on a glass bottom culture dish (15\(\text{mm}\), 180 \(\mu m\) thick).

\subsubsection{Ground truth estimation}

\begin{figure}[ht!]
\centering
\includegraphics{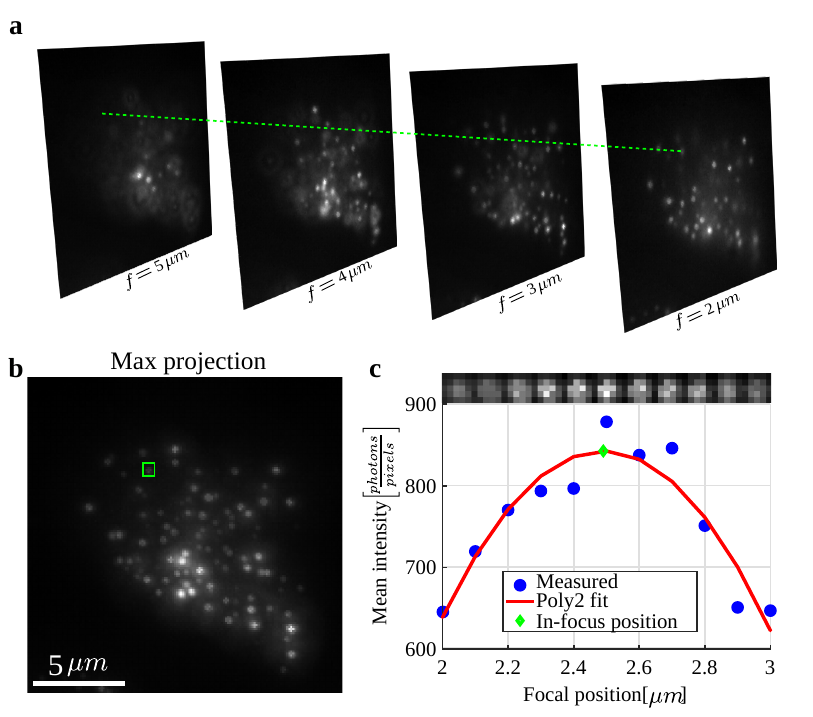}
\caption{Experimental ground truth approximation. (a) A focal sweep is performed with an unmodulated imaging path. (b) Max projection of the focal sweep, showing the density of labelled telomeres in the U2OS cell. (c) Axial fit of the mean intensity to determine the in-focus position of an emitter.}
\label{figs:gt-approx}
\end{figure}

In fixed cell experiments, the experimental ground truth 3D positions were approximated \textit{via} axial scanning with the unmodulated PSF (Fig. \ref{figs:gt-approx}). The scan consisted of 100 nm steps over a range of 4-5 \(\mu m\). The resulting z-stack was then processed in the following manner: first, detection and lateral position estimation were performed with ThunderSTORM \cite{ovesny2014thunderstorm}. Next, the in-focus position of emitters was estimated by fitting a second order polynomial to the mean intensity across focal slices. The mean intensity was calculated as the mean of number of counts in the central \(5\times5\) pixels of each detected PSF. The emitter axial position was obtained by correcting the detected in-focus position with a factor of 0.8 accounting for refractive index mismatch. Since VIPR \cite{ferdman2020vipr} accounts for the 3D wobble of the modulated PSFs, the final required correction was a global lateral shift between the in-focus PSF and the chosen modulated-PSF center in the phase retrieval step.

\begin{figure*}[htp!]
\centering
\includegraphics{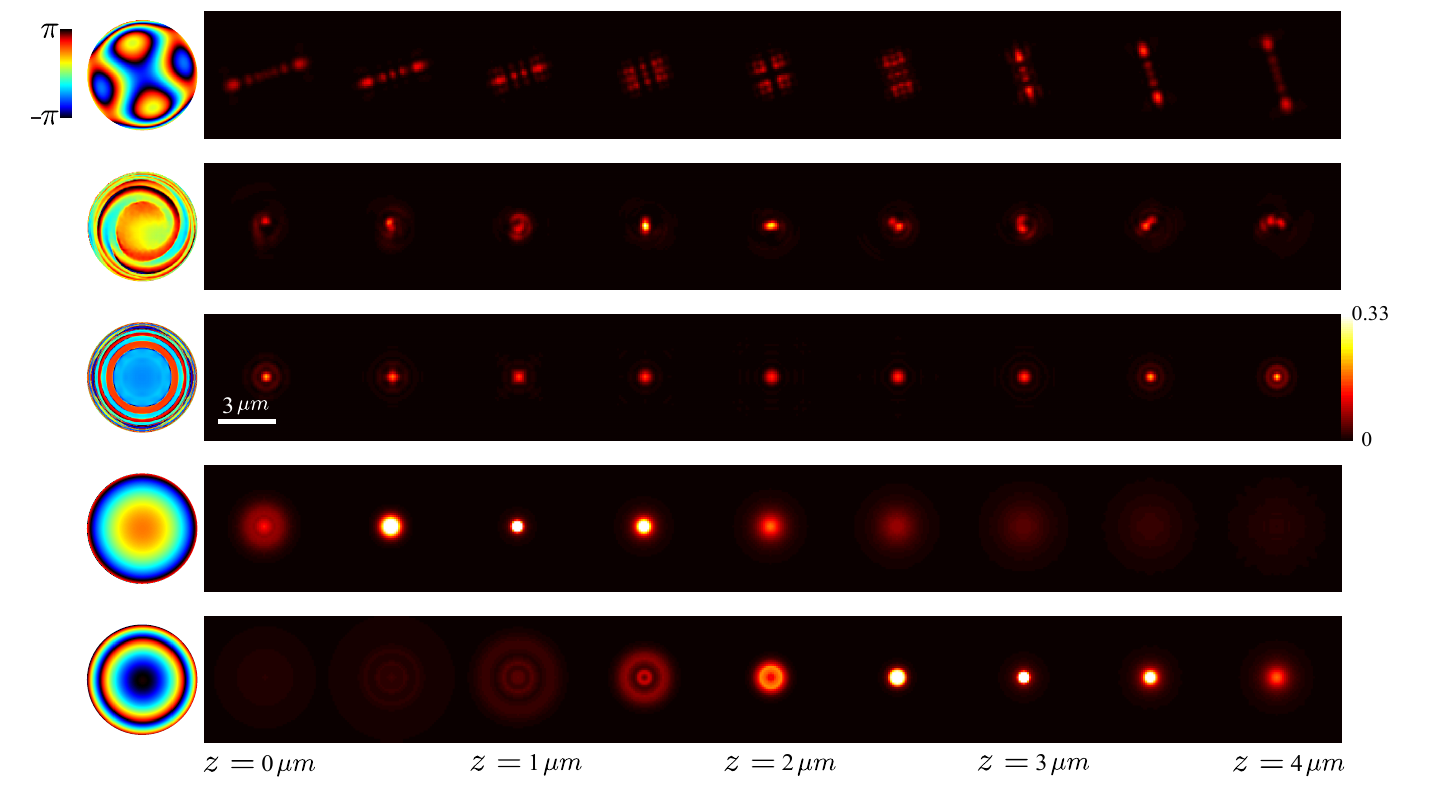}
\caption{PSFs for single and dual channel comparisons. Phase masks which were used in the single-channel vs. dual-channel comparison: (top to bottom) Tetrapod, end-to-end encoding for single-channel, EDOF, and biplane.} 
\label{figs:single-vs-dual-psfs}
\end{figure*}

\begin{figure}[htp!]
\centering
\includegraphics{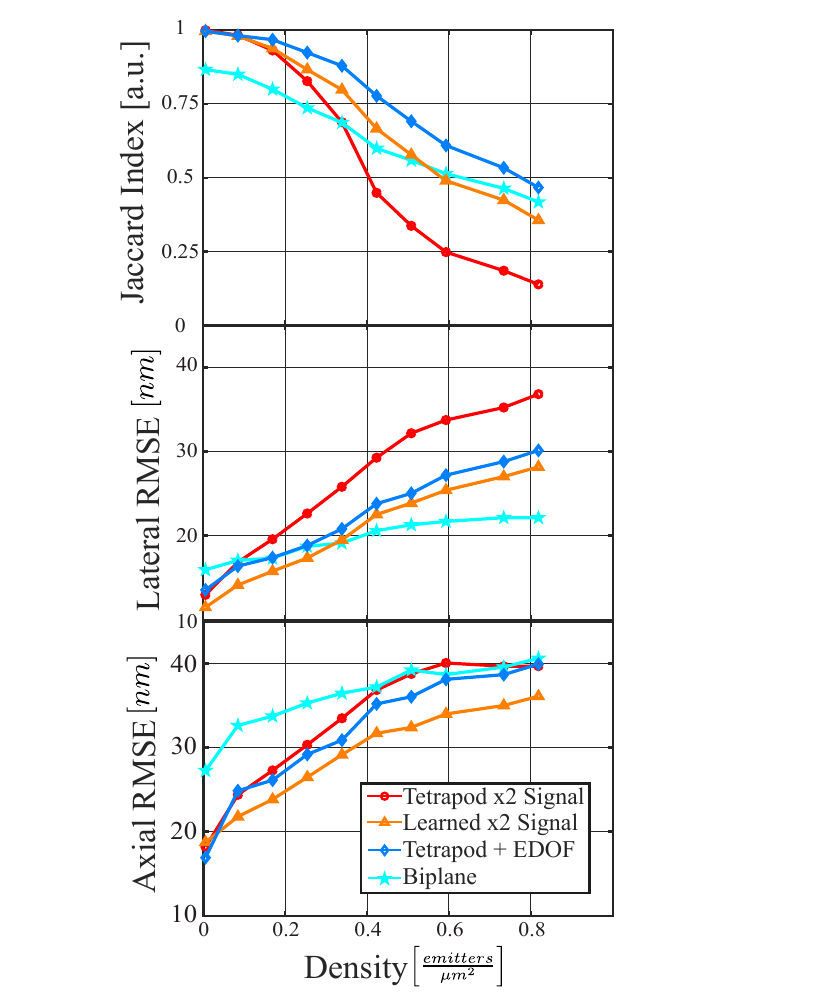}
\caption{Single-channel vs. dual-channel systems. Detection (left) and localization precision (lateral \textbackslash axial) over a range of simulated density of sources. Emitters were simulated with \(\approx\)15K signal photons per emitter and \(\approx\)500 background photons per pixel. Each data point is an average of n = 100 simulated images. Average standard deviation in Jaccard index was \(\approx\)5\% and in precision was \(\approx\)3 nm.} 
\label{figs:single-vs-dual-plots}
\end{figure}

\begin{figure}[htp!]
\centering
\includegraphics{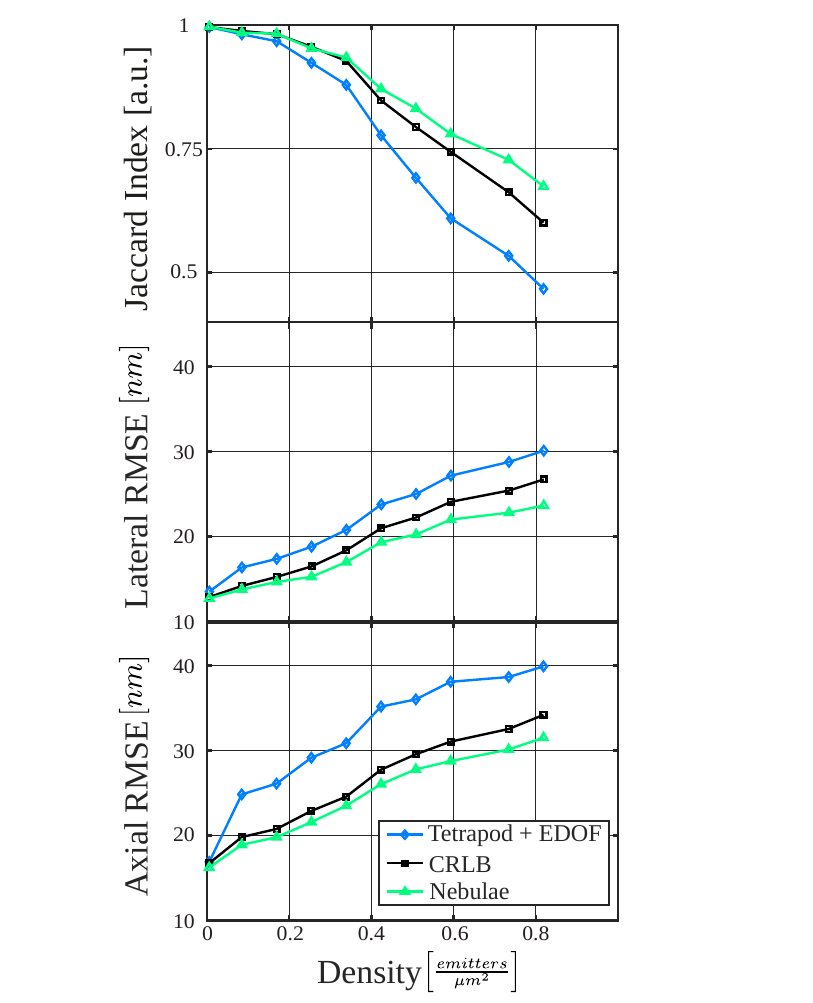}
\caption{Performance as function of density for the three proposed PSF pairs. The methods are tested in detection (left) and localization precision (lateral \textbackslash axial RMSE). Emitters were simulated with \(\approx\)15K signal photons per emitter and \(\approx\)500 background photons per pixel. Each data point is an average of n = 100 simulated images. Average standard deviation in Jaccard index was \(\approx\)5\% and in precision was \(\approx\)3 nm.} 
\label{figs:3pair-sim}
\end{figure}

\subsection{Additional simulation results} \label{supp-add-sim}

In this section we present further numerical simulation results which support conclusions from the main text and the choice of the PSF pair. The first result presented in Figs. \ref{figs:single-vs-dual-psfs} and  \ref{figs:single-vs-dual-plots} shows a numerical comparison between single-channel and dual-channel setups in terms of their detection (measured by the Jaccard index) and the average precision (measured by the lateral\textbackslash axial RMSE).  We compare the Tetrapod-EDOF (blue) pair to the commonly used biplane (cyan) method \cite{juette2008three,ram2008high} and to two single-channel approaches with double signal, namely the Tetrapod PSF (red) and the single channel end-to-end optimized phase mask (orange) adopted from DeepSTORM3D \cite{nehme2020deepstorm3d}. The numerical results show that the Tetrapod-EDOF pair is the best in detection. In terms of the lateral RMSE in high densities, the biplane approach is better as the in-focus PSF is more photon efficient than the EDOF. The axial RMSE result shows that the proposed pair is surpassed only by the end-to-end encoding of a single channel. This is likely because the axial position is mainly encoded in the Tetrapod path, thus is limited to the axial localization performance of the single channel Tetrapod at high densities. These results reinforce the decision to explore other solutions which mutually encode all parameters in both channels, and are optimal for detection and localization.

The second result in Fig. \ref{figs:3pair-sim} is a comparison between the three proposed PSF pairs in this manuscript. Both the detection and the average precision support our claim that the Nebulae PSFs (green) are better than the CRLB (black) and Tetrapod-EDOF pairs (blue). A similar conclusion was drawn from the experimental results in fixed cells, which ultimately supports our decision to use the Nebulae PSFs for live cell tracking. Comparing all the tested metrics, we can see that at every density, the Nebulae PSFs are undoubtedly the best choice out of the three.

\subsection{Additional experimental results} \label{supp-add-exp}

\begin{figure}[h!]
\includegraphics{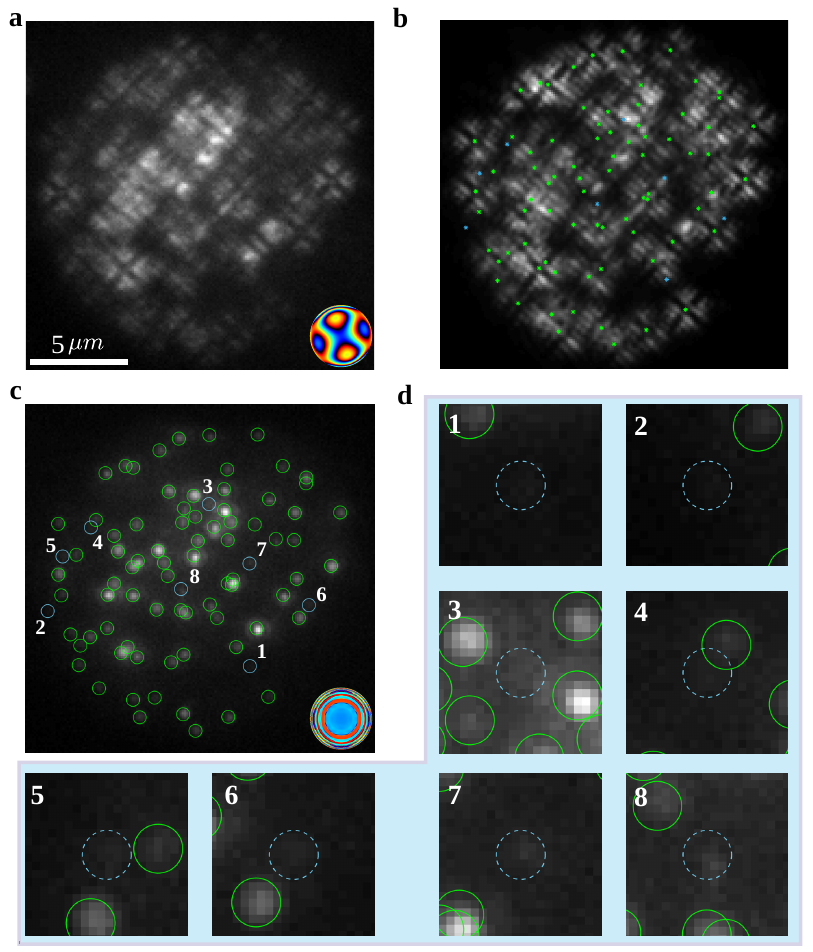}
\caption{Experimental false negatives for the Tetrapod-EDOF pair.(a) U2OS cell experimental snapshot with the Tetrapod PSF (Fig. \ref{fig:tp-vs-tp-edof}). (b) Reconstructed image by rendering the positions recovered by the net with the Tetrapod PSF. Asterisks mark true (green) and false (blue) positives. (c) Paired experimental EDOF snapshot. (d) Zoom-ins on undetected emitters (false positives).}
\label{figs:oracle-miss}
\end{figure}

This section presents more experimental results in fixed cell data. Figure \ref{figs:oracle-miss} explores the false negatives presented in \ref{fig:tp-vs-tp-edof}. All of the experimentally undetected points were with a very low signal. While the EDOF performs well in 2D, it is not as signal efficient as the in-focus unmodulated PSF. Thus, emitters which are slightly above the noise limit (without a phase mask) can be detected in the axial scan but are invisible for the EDOF and Tetrapod PSFs. This was improved in the subsequent PSF-pairs which complement each other more efficiently. 

\begin{figure*}[htp!]
\centering
\includegraphics{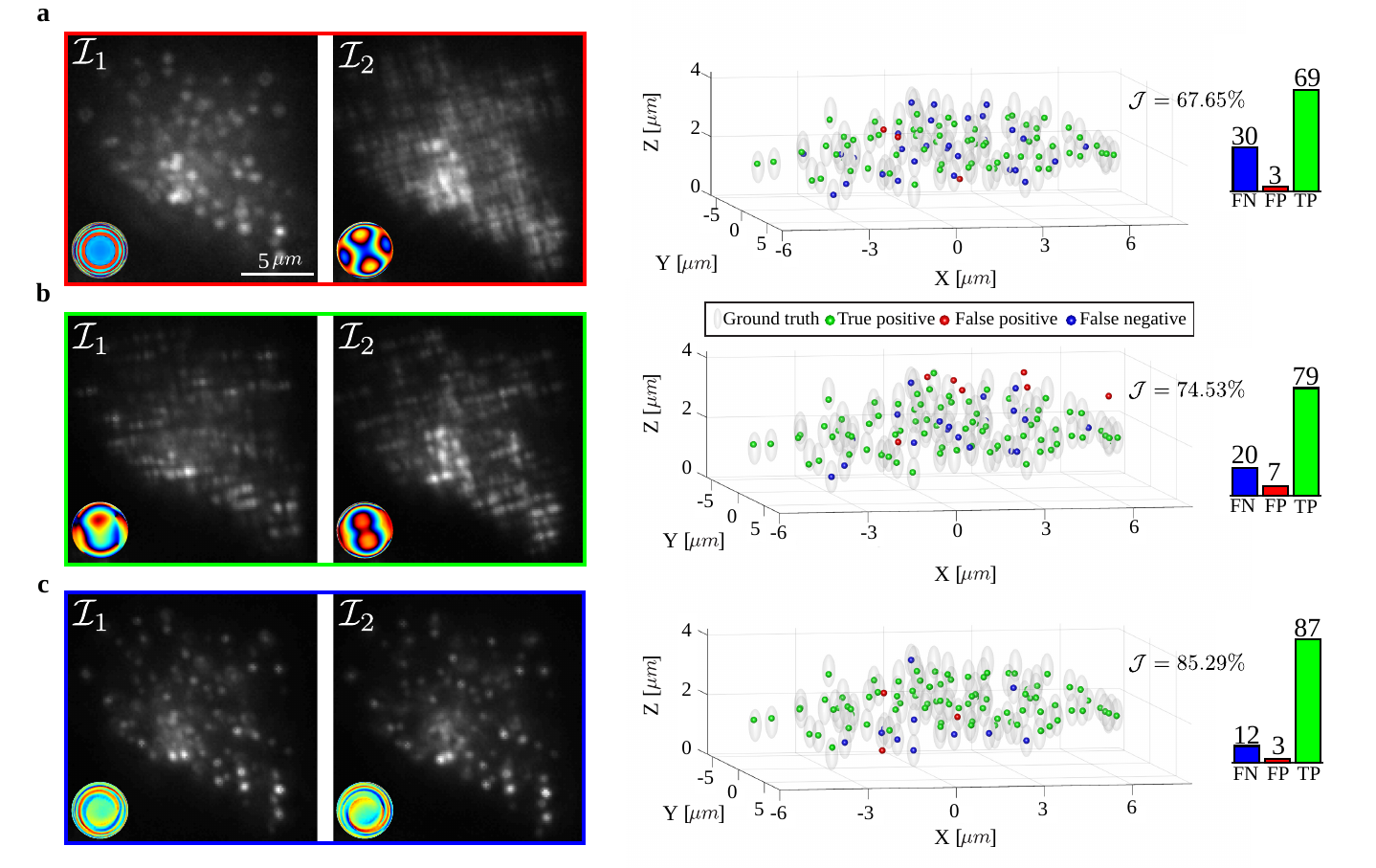}
\caption{Experimental measurement of fixed U2OS cells with fluorescently labelled telomeres. Example images showing the two proposed mask image pairs and subsequent 3D reconstruction plotted over the approximated ground truth: (a) Tetrapod-EDOF pair with \(\mathcal{J}=67.65\%\), (b) CRLB pair with \(\mathcal{J}=74.53\%\), and (c) Nebulae PSFs with \(\mathcal{J}=85.29\%\).}
\label{figs:3pairs-main-crlb}
\end{figure*}

\begin{figure*}[htp!]
\centering
\includegraphics{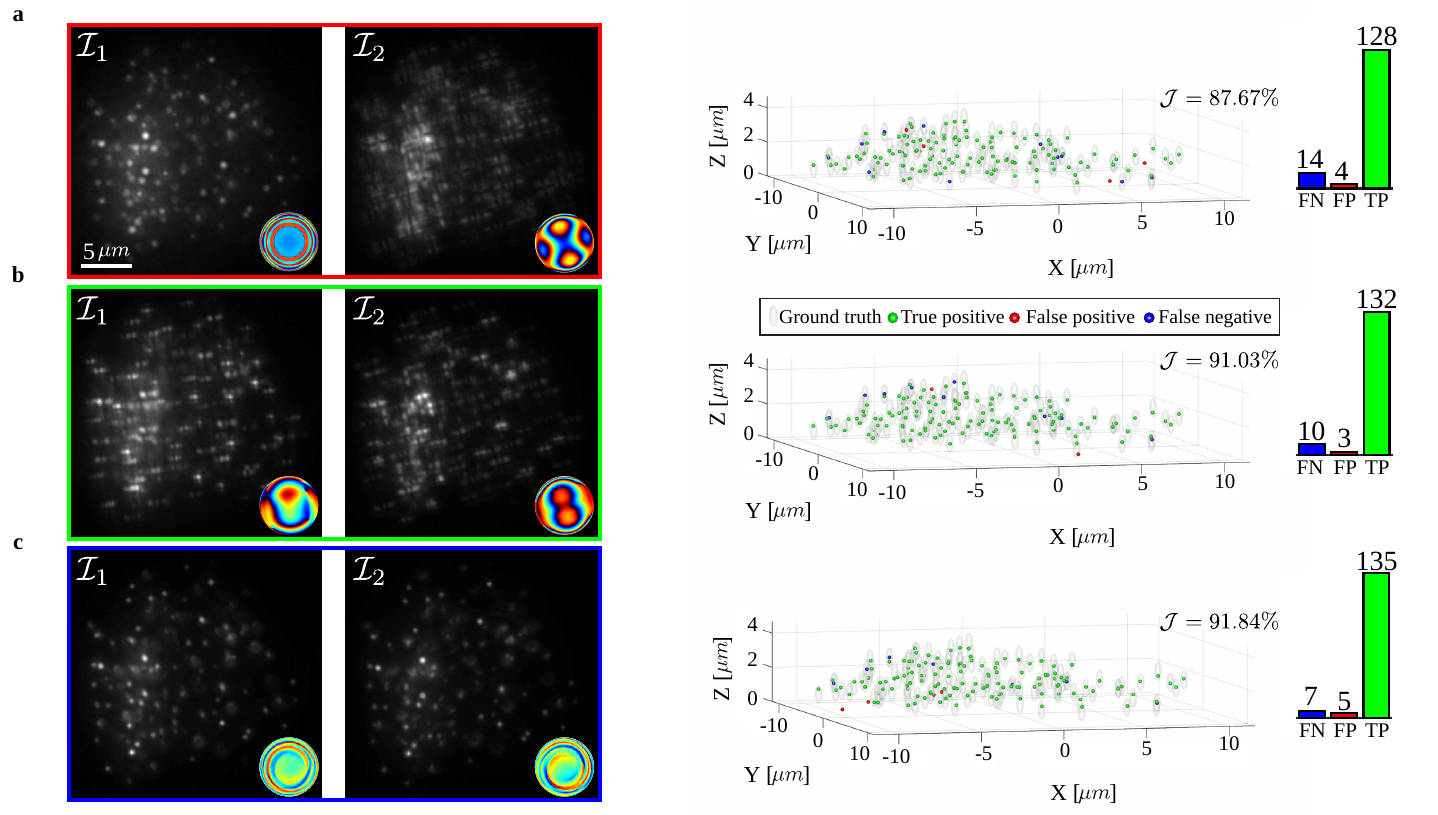}
\caption{Experimental measurement of fixed U2OS cells with fluorescently labelled telomeres. Example images showing the two proposed mask image pairs and subsequent 3D reconstruction plotted over the approximated ground truth: (a) Tetrapod-EDOF pair with \(\mathcal{J}=87.67\%\), (b) CRLB pair with \(\mathcal{J}=91.03\%\), and (c) Nebulae PSFs with \(\mathcal{J}=91.84\%\).}
\label{figs:3pairs-exp2}
\end{figure*}

To validate our conclusions from simulation regarding the Nebulae PSFs being the optimal pair, we have shown in Fig. \ref{figs:3pairs-main-crlb} that the Nebulae PSFs are outperform the Tetrapod-EDOF pair. For completeness, we show in Fig. \ref{figs:3pairs-main-crlb} the results including the CRLB-pair for the same cell. As predicted in simulations, the CRLB pair performs slightly worse than the Nebulae PSFs but better than the Tetrapod-EDOF pair. To verify reproducibility, we present in Fig. \ref{figs:3pairs-exp2} similar experimental results for a bigger cell, which exhibits a staggering number of 142 emitters. The reconstruction results are improved for all PSF pairs as this cell experiences less overlaps, yet, they are consistent with the previous conclusions on PSF-pair performance.

\bibliographystyle{ieeetr}
\normalsize{\bibliography{main_text}}

\end{document}